\documentclass[letterpaper]{article} 

\usepackage[usenames,dvipsnames]{color}
\usepackage{bm}
\usepackage{graphicx,fancyhdr,lscape,dsfont,amsmath,amssymb} 
\usepackage{color, amsthm}
\usepackage{longtable}
\usepackage{psfrag}
\usepackage{subcaption}
\usepackage{tabularx}
\usepackage{boldline,multirow}
\usepackage{stmaryrd}
\usepackage{appendix}
\usepackage{soul}
\usepackage{cancel}
\usepackage{chemmacros}
\usepackage{ulem} 
\usepackage{sidecap}
\usepackage{appendix}
\usepackage{booktabs}
\usepackage{lmodern}

\sidecaptionvpos{figure}{c}

\allowdisplaybreaks

\newcommand{\vect}[1]{\vec{#1}} 
\newcommand{\tensor}[1]{  {\bm {#1}} } 

\newcommand{\trace}[1]{ {\rm tr} \left[ \, {#1} \, \right] }

\newcommand{\deviatoric}[1]{ {\rm dev} \left[ \, {#1} \, \right] }
\newcommand{\divergence}[1]{ {\rm div} \left[ \, {#1} \, \right] }
\newcommand{\gradient}[1]{ {\rm \nabla} \left[ \, {#1} \, \right] }

\newcommand{\diffusivity}{\mbox{${\rm D} \mskip-8mu  | \,$}}

\newcommand{\permittivity}{\mbox{${\varepsilon} \mskip-7mu  | \,$}}

\renewcommand{\em}[1]{\it{#1}}
\newcommand{\ra}[1]{\renewcommand{\arraystretch}{#1}}

\graphicspath{{./}{./Figures/}}

\usepackage[utf8]{inputenc}
\DeclareUnicodeCharacter{2212}{-}

\begin{document}

\title{Quantitative investigation of the influence of electrode morphology in the electro-chemo-mechanical response of Li-ion batteries.} 
\author{Magri M.$^{1}$,  Boz B.$^{2}$, Cabras L.$^{2}$,  Salvadori A.$^{2}$
 \\
\begin{small}
	$^{1}$  Fundacion IMDEA Materiales, C/Eric Kandel 2, 28906, Getafe, Madrid, Spain,
\end{small}
\\
\begin{small}
$^{2}$ {\sf m$^4$lab}, DIMI, Universit{\`a} di Brescia, via Branze 38, 25123 Brescia, Italy 
 \end{small}
}

\maketitle

\begin{abstract}
In Li-ion batteries the electrochemical potential drives the redox reactions occurring at the interface between electrolyte and storage material, typically active particles for porous electrodes, allowing Li ions intercalation/extraction from electrodes. The limiting factor of the process is not solely related to the electrochemical affinity, but also to structural features as the morphology of the electrodes. In this note, the relevance of the electrodes architecture is investigated numerically, in terms of the electro-chemo-mechanical response of a battery cell with homogeneous electrodes and conventional liquid electrolyte. 
 A suitable design of the electrodes induces major increments of battery efficiency.

\end{abstract}

\section{Introduction}

Because of their high working voltage and large power density, Li-ion batteries (LiBs henceforth) are suitable for powering electric and hybrid vehicles (HEVs) \cite{harper2019recycling}. Green mobility, though, demands for LiBs advancement in safety, life time, and capacity preservation with cycling. To improve the energy density of LiBs, two main methods could be suggested that change their intrinsic properties: either developing novel battery chemistry or architecting electrodes. Countless efforts have been spending into improving the existing cathode materials \cite{lain2019design,wang2018toward,kremer2020manufacturing}. Less has been done in terms of electrode structural design \cite{kuang2019thick,mei2019effect}.

\begin{figure}[htbp]
\centering
 \includegraphics[width=16.5cm]{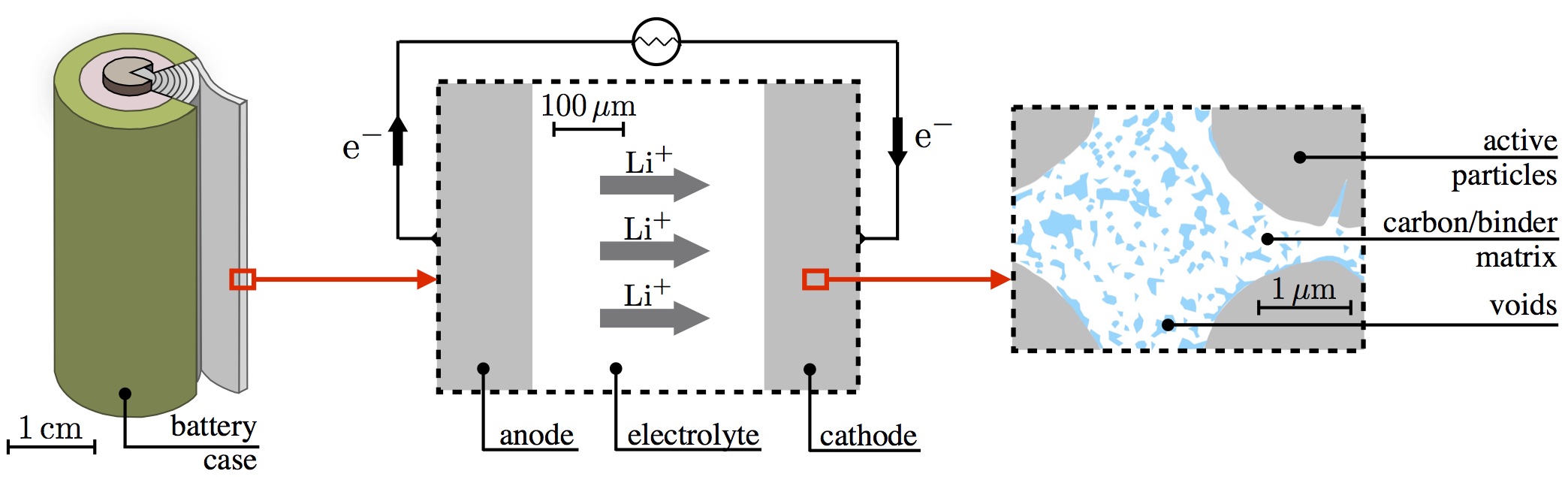}
\caption{\em{Schematic representation of a typical electrochemical cell, the characteristic length at different scales, and the paradigmatic microstructure of a porous electrode.}}
\label{fig:battery_scheme}
\end{figure}

As depicted in Fig. \ref{fig:battery_scheme}, conventional LiBs consists of one anode, one cathode, and a liquid
electrolyte with a separator, which is a porous polyethylene or polypropylene material used to create ionic contact while maintaining electronic insulation between the two electrodes and current collectors. During battery discharging, the electrochemical affinity triggers the motion of Li ions from the anode to the cathode across the electrolyte, as well as the flow of electrons, with the same direction, through the external circuit; vice versa during charging. 
In conventional systems, the porosity of the carbon/binder matrix is filled with liquid electrolyte, which penetrates from the separator up to the particle surface. 
The carbon/binder matrix, which is electrochemically inert, increases the overall electrical conductivity of the electrode and provides structural integrity. Electrodes are made porous to increase the surface over which the electrochemical charge transfer reaction can occur, thus increasing the efficiency of batteries \cite{AifBook}.


Although the scenario depicted in Fig. \ref{fig:battery_scheme} is quite simple, modeling LiBs is challenging, as their behavior is intrinsically multi-physics and multi-scale  \cite{FrancoRCS2013, GrazioliEtAlCM2016, liu2019simultaneous}. Modeling, therefore, requires complex systems of coupled nonlinear time-dependent partial differential equations, capable of governing electrochemical-thermal-mechanical processes over a range of spatio-temporal scales. The broad nature of the phenomena involved (mechanical, electrical, electrochemical, and thermal) and the interactions among them lead to complex mathematics with a very high number of unknown fields (displacements, electric potential, concentrations, temperature). Macroscale modelling is more suitable to evaluate the overall battery performance \cite{arunachalam2019full,battiato2019theory} but their phenomenological nature may prevent predictive ability in tailoring intrinsic microstructural properties. 
Indeed, the most relevant phenomena take place at the characteristic length scale of the electrode compound, which can be several order of magnitude smaller then the battery size \cite{higa2017comparing}. Gaining an understanding of the role of these processes by theoretical and numerical studies can lead designing better performing batteries \cite{lu20203d,kim2018multiphysics}. 



Computational simulations, based on rigorous theoretical modeling and coupled to validation and quantification of the uncertainties, have the potential to enhance batteries’ performances through new shapes and new materials \cite{zhu2021numerical,sonwane2021coupling}.  Numerical and experimental studies, in fact, enabled to grasp a much more profound understanding of the multiphysics processes that occur during charge and discharge. Moreover, due to the high costs and time required by experimental studies, numerical simulations became more widespread in the battery community, increasing in complexity and capabilities. 

\begin{figure}[htbp]
\centering
 \includegraphics[width=16.5cm]{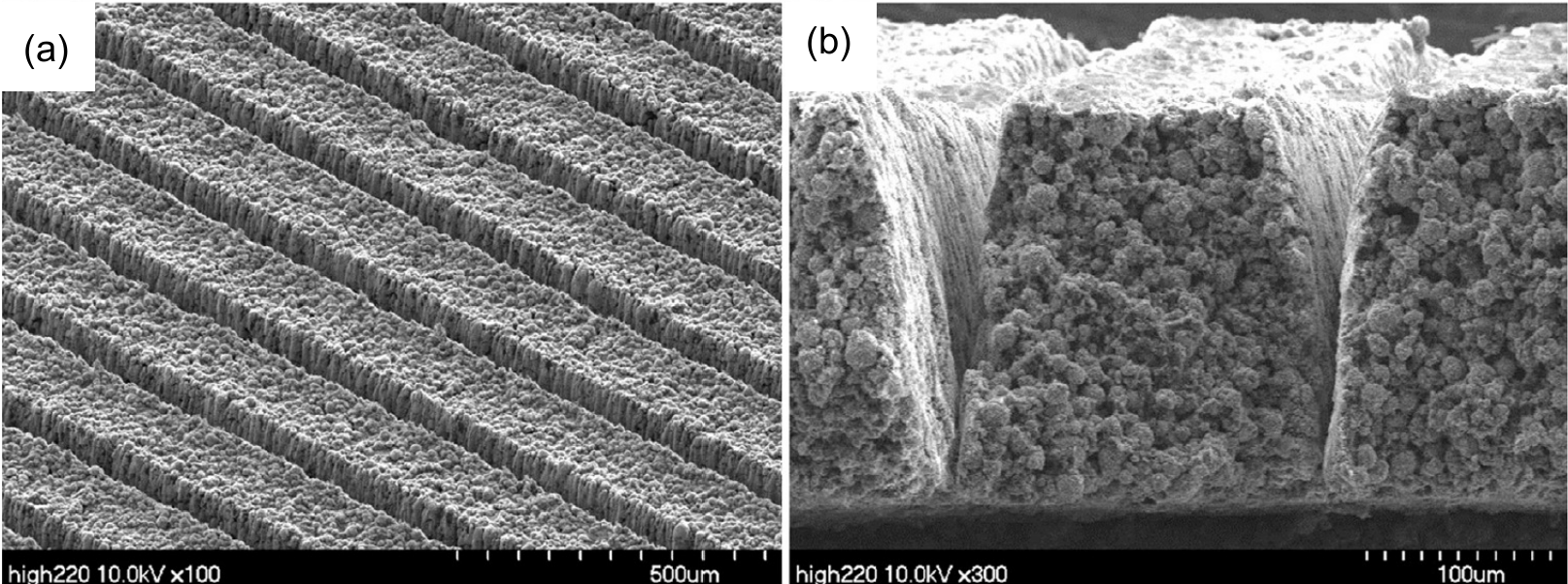}
\caption{\em{SEM images of structured electrode a) structured electrode with high porosity. b) the cross section of structured electrode. Reprinted and adapted with permission from reference \cite{ParkJIEC2019}}}
\label{fig:battery_structured}
\end{figure}

Substantial experimental efforts on lithium ion batteries have been carried out to meet the high energy/power density requirements in terms of new battery configurations and material chemistry \cite{ZhangAEM2021}. Despite the progresses in advanced materials, the active components generally behave differently from the theoretical expectations, showing significant energy loss and capacity failures. To overcome these issues, engineering the electrodes by increasing exposed transport channels or tailoring the electrochemical properties of the electrodes are suggested as promising solutions \cite{YanAEM2018}.
Architecting thick electrodes has been investigated broadly with several experimental techniques such as: metal foams and carbon textile based thick cathodes \cite{JiNL2012}, 3D embroidered aluminum current collectors \cite{AguiloRSC2016}, formation of vertical open channels till current collector \cite{EvanoffAM2012,SanderNE2016}, laser structuring to create through holes till current collector \cite{Park2010,MangangJPS2016,Pfleging2018,TsudaEA2019}. The most promising experimental studies display the idea of our theoretical work.
Park and colleagues designed a lithium nickel manganese cobalt oxide (NMC) electrode with the thickness of 175 $\mu$m with 26\% porosity \cite{ParkJIEC2019}. The laser structuring process was used to turn the flat electrode film into a three-dimensional patterned microstructure, see Fig. \ref{fig:battery_structured}. The specific energy of the thick and dense electrodes increased with porosity, while improving or retaining the power density at high current rates owing to the enlarged electrode surface area. Tsuda and coworkers synthesized ultra-thick lithium iron phosphate ($\rm LiFePO_4$) holing cathodes by pico-second pulsed lasers, with hole diameters of approximately 20-30 $\mu$m \cite{TsudaEA2019}. Making holes through the electrode decreased the charge transfer resistance, since the available area for Li ions to diffuse increased. Ultimately, increased surface area and reduction in tortuosity lead to a gain in the overall battery performance. Lim and coworkers designed 3D conical-shaped microelectrodes based on experimental evidences \cite{LimJES2014}. A combined electrochemical and chemo-mechanical model was proposed to simulate a three- dimensional $\rm LiCoO_2$-based conical structure. It was found that the conical structure mechanically relaxes the substrate, thus reducing the possibility of mechanical failure-induced capacity loss. Pfleging’s extensive review \cite{Pfleging2018} on laser electrode processing for the improvement of lithium ion battery systems is recommendable.

In this work, we aim at tailoring electrode configurations via computational simulations, seeking for optimal functioning of conventional LiBs. We aim at {\em{optimizing the electrode design}} to alleviate the limiting factors of battery power delivery, which can be categorized as: i) the rate of diffusion of Li cations into cathode or anode particles, ii) the rate and distance of ionic transport in the electrolyte and between the electrode particles and iii) the electrical conductivity of cathode materials.

In this paper, the influence of electrodes morphology on the response of the battery is evaluated numerically. After nucleating the governing equations of a multiscale-compatible modeling approach in section \ref{sec:mat_mod}, we considered a battery cell made up of homogeneous electrodes and a liquid electrolyte and devised a unit cell for an electrode architecture as a function of a porosity parameter in section \ref{sec:pb_def}. Outcomes have been largely discussed afterwards, showing how tailoring the electrode configuration could ideally boost the battery efficiency from 3\% to almost 100\%. Limits and further developments of the current work conclude the paper.


\section{Mathematical model} \label{sec:mat_mod}

\subsection{Electrolyte} \label{electrolyte}
The battery cell is considered to have 1 M LiPF$_6$ liquid electrolyte, which is a solution of a binary salt in carbonate solvent mixture. The lithium salt is denoted for simplicity as LiX. The electrolyte is then characterized by the presence of ionic species Li$^{+}$ and X$^{-}$ after the complete decomposition of the salt.  The electrolyte is modeled following the approach of \cite{SalvadoriEtAlJPS2015}, which assumes the transport of dissolved ions as driven by diffusion and migration in a non convecting medium. Differently from widespread models of electrolytes (see \cite{NewmanInScrosatiBook2002} for example), the kinetics of positive and negative ions is modeled independently, without neglecting charge separations and the related electro-magnetic interactions. As pursued in \cite{SalvadoriEtAlJPS2015}, the electro-quasi-statics approximation of the Maxwell's equations is considered, by neglecting the time derivative of the magnetic field. For the sake of brevity we address the reader to \cite{SalvadoriEtAlIJSS2015,SalvadoriEtAlJMPS2013,SalvadoriGrazioliBookCh2015} for details. The balance laws in the electrolyte domain $V_e$ include
\begin{itemize}

\item Conservation of moving ions

\begin{subequations} \label{eq:el_mass_balance}
\begin{gather}
\frac{\partial c_{\text{Li}^+}}{\partial t}  +\divergence{ \vect{h}_{\text{Li}^+}} = 0 \, , \\
\frac{\partial c_{\text{X}^-}}{\partial t}  + \divergence{\vect{h}_{\text{X}^-}} = 0  \, ,
\end{gather}
\end{subequations}

\noindent
where $c_\alpha$ is the molarity (i.e. the number of moles per unit volume) of the specie $\alpha=\text{Li}^+$, $\text{X}^-$; $\vect{h}_\alpha$ is the mass flux in terms of moles, i.e. the number of moles of species $\alpha$ measured per unit area per unit time. 

\item Maxwell's equations for electro-quasi-statics

\begin{subequations} \label{eq:el_maxwell}
\begin{gather}
\divergence{\frac{\partial \vect{D}_e}{\partial t} + \vect{i}_e} = 0  \, ,\\
\vect{E}_e = -\gradient{\phi_e} \, ,
\end{gather}
\end{subequations}

where $\vect{D}_e$, $\vect{i}_e$, ${E}_e$, and $\phi_e$ refer to the electric displacement, current density, electric field, and electric potential in the electrolyte respectively.

\item Balance of linear and angular momentum

\begin{subequations} \label{eq:el_mechanics}
\begin{gather}
\divergence{\tensor{\sigma}_e} + \vect{b} = \vect{0} \, , \label{eq:el_mechanicsa} \\
\tensor{\sigma}_e = \tensor{\sigma}_e^T \, , 
\end{gather}
\end{subequations}

where $\tensor{\sigma}_e$ is the Cauchy stress tensor, and $\vect{b}$ is the vector of body forces, i.e. force per unit volume. Inertial effect are neglected in equation \eqref{eq:el_mechanicsa}.

\end{itemize}

The connection among balance equations \eqref{eq:el_mass_balance}-\eqref{eq:el_mechanics} is provided by the constitutive definition of $\vect{h}_\alpha$, $\vect{D}_e$, $\vect{i}_e$, and $\tensor{\sigma}_e$ in terms of the independent variables $c_\alpha$, $\phi_e$, and displacement vector $\vect{u}_e$. The flux of ionic species is constitutively described by the Nernst-Planck formulation, extended to take into account the saturation of electrolyte species as in \cite{SalvadoriEtAlJPS2015b}
\begin{subequations} \label{eq:ionic_fluxes}
\begin{gather}
\vect{h}_{\text{Li}^+} = - \diffusivity_{\text{Li}^+} \, \gradient{c_{\text{Li}^+}} - \frac{\diffusivity_{\text{Li}^+} \, F}{R \, T} \, c_{\text{Li}^+} \, \left( 1 - 2 \, \frac{c_{\text{Li}^+}}{c^{max}} \right) \, \gradient{\phi_e} \, , \\
\nonumber
\vect{h}_{\text{X}^-} = - \diffusivity_{\text{X}^-} \, \gradient{c_{\text{X}^-}} + \frac{\diffusivity_{\text{X}^-} \, F}{R \, T} \, c_{\text{X}^-} \, \left( 1 - 2 \, \frac{c_{\text{X}^-}}{c^{max}} \right) \, \gradient{\phi_e} \, .
\end{gather}
\end{subequations}
In equations \eqref{eq:ionic_fluxes},  $\diffusivity_{\alpha}$ is the diffusion coefficient of species $\alpha$; $c^{max}$ is the ionic saturation limit; symbols $F$, $R$, and $T$ refer to Faraday constant, universal gas constant, and absolute temperature, respectively. 
The current density comes out from Faraday's law as 

\begin{equation}
\vect{i}_e  = F \, \left( \vect{h}_{\text{Li}^+} -  \vect{h}_{\text{X}^-} \right) \, .
\end{equation}

\noindent
Assuming the electrolyte as an isotropic, linear, homogeneous dielectric, the constitutive definition of the electric displacement reads

\begin{equation} \label{eq:const_def_disp_el}
\vect{D} = \permittivity \, \vect{E} \, ,
\end{equation}

\noindent
where $\permittivity = \permittivity_r \, \permittivity_0$ is the electrolyte permittivity, which is given, as usual, relative to that of vacuum (denoted with $\permittivity_0$) as a relative permittivity $\permittivity_r$.
From the mechanical point of view, the electrolyte is assumed as an isotropic-linear-elastic medium without splitting of deformation, viewing the latter as independent upon the ionic concentration; accordingly the stress tensor relates to the elastic strain tensor $\tensor{\varepsilon}_e$ through the following relationship
\begin{equation}
\tensor{\sigma}_e = K_{e} \,  \trace{\tensor{\varepsilon}_e} \mathds{1} + 2 \, G_e \, \deviatoric{\tensor{\varepsilon}_e} \, ,
\end{equation}
where $K_e$ is the bulk modulus and $G_e$ is the shear modulus. The elastic strain tensor is function of $\vect{u}_e$ through the formula
\begin{equation}
\tensor{\varepsilon}_e = \frac{1}{2} \left( \gradient{\vect{u_e}} + \gradient{\vect{u_e}}^T \right) \, .
\end{equation}
as in small-strains theories.

\subsection{Intercalating electrodes} \label{int_electrodes}

Electrodes are here assumed as homogeneous media, characterized by the presence of intercalated Li ions and moving electrons. Li ions in active materials are screened by the mobile electrons, which accompany lithium when moves from one interstitial site to the other \cite{Danilovetal2011}.  Therefore, the charge of Li cation after intercalation into active materials is instantaneously wiped out by the transport of electrons over the current collectors toward the particle surface. The active storage materials are thus idealized as interstitial solid solution containing dissolved Li and electrons, which are free to move. Because active materials have much higher electronic conductivity than the electrolyte, the electromagnetic problem will be considered in the electro-static approximation.  According to the framework developed in \cite{SalvadoriEtAlIJSS2015}, the governing equations for the electrodes can be summarized as follows.

\begin{itemize}
\item Conservation of dissolved lithium
\begin{equation} \label{eq:a_mass_balance}
\frac{\partial c_{\text{Li}}}{\partial t} + \divergence{\vect{h}_\text{Li}} = 0 \, 
\; ,
\end{equation}
where $c_\text{Li}$ is the molarity of Li and $\vect{h}_\text{Li}$ is its flux vector. 

\item Charge conservation
\begin{equation} \label{eq:steady_ch_cons}
\divergence{\vect{i}_a} = 0 \, ,
\end{equation}
where $\vect{i}_a$ refers to the current density in the electrodes. 

\item Balance of linear and angular momentum
\begin{subequations} \label{eq:elct_bal_mom}
\begin{gather}
\divergence{\tensor{\sigma}_a} + \vect{b} = \vect{0} \, , \\
\tensor{\sigma}_a = \tensor{\sigma}_a^T \, , 
\end{gather}
\end{subequations} 
with obvious meaning of symbols $\tensor{\sigma}_a$ and $\vect{b}$.

\end{itemize}
 
Constitutive equations are derived from thermodynamic principles, as carried out in \cite{SalvadoriEtAlJMPS2018}, and are given in terms of lithium concentration $c_\text{Li}$, electric potential $\phi_a$, and displacement vector $\vect{u}_a$. To account for the effects of lithiation, the strain tensor $\tensor{\varepsilon}_a = \frac{1}{2} ( \gradient{\vect{u_a}} + \gradient{\vect{u_a}}^T )$ is additively decomposed in two contributions
\begin{equation} \label{eq:a_strain_decomposition}
\tensor{\varepsilon}_a = \tensor{\varepsilon}^{el}_a + \tensor{\varepsilon}^{ch}_a \, ,
\end{equation}
where $\tensor{\varepsilon}^{el}_a$ is the elastic part of the deformation, while the chemical strain $\tensor{\varepsilon}^{ch}_a$ is the volumetric deformation of the electrode lattice associated with lithium intercalation as
\begin{equation} \label{eq:chemical_strain_tensor}
\tensor{\varepsilon}^{ch}_a = \omega_\text{Li} \left( c_\text{Li} - c_\text{Li}^0 \right) \, \mathds{1} \, .
\end{equation}
In formula \eqref{eq:chemical_strain_tensor}, $\omega_\text{Li}$ is the coefficient of chemical expansion, i.e. one-third of Li partial molar volume at given temperature, while $\mathds{1}$ denotes the identity tensor. Owing to formula \eqref{eq:a_strain_decomposition}, the stress tensor is defined as

\begin{equation} \label{eq:stress_def}
\tensor{\sigma}_a = K_{a} \,  \trace{\tensor{\varepsilon}_a - \tensor{\varepsilon}^{ch}_a} \mathds{1} + 2 \, G_a \, \deviatoric{\tensor{\varepsilon}_a} \, ,
\end{equation}

\noindent
with obvious meaning of constants $K_a$ and $G_a$. Constitutive law \eqref{eq:stress_def} is extremely simple if compared with recent literature as \cite{Ganser2019,dileoetalJMPS2014,DiLeoIJSS2015a}. In fact, we focus here on the influence of electrode morphology on the charge/discharge without profoundly investigating the mechanical performance of batteries. Nonetheless, as it will be elaborated later on in the paper, mechanical effects cannot be neglected.

In a thermodynamic consistent theory, the definition of the Li flux in Eq. \eqref{eq:a_mass_balance} is derived from the generalized Fick's law. For the sake of simplicity, the effect of complicate phenomena, such as phase-segregation, are not considered in this study. Accordingly, the chemical potential of dissolved lithium has the usual expression for single-phase intercalating electrodes
\begin{equation} \label{eq:chem_pot}
\mu_\text{Li} = \mu_\text{Li}^0 + R \, T \, \text{ln} \left( \frac{c_\text{Li}}{c_\text{Li}^{max} - c_\text{Li} } \right) - \omega_\text{Li} \, \trace{\tensor{\sigma}_a} \, ,
\end{equation}
with $\mu_\text{Li}^0$ denoting the reference chemical potential. Therefore, the flux of dissolved lithium in the electrodes obeys the coupled constitutive equation
\begin{equation} \label{eq:Li_flux}
\vect{h}_{\text{Li}} = -\diffusivity_\text{Li} \, \gradient{c_\text{Li}} + \frac{\diffusivity_\text{Li} \, \omega_{Li} }{ R \, T } \, c_\text{Li} \, \left( \frac{c_\text{Li}^{max} - c_\text{Li} }{c_\text{Li}^{max}} \right) \, \gradient{\trace{\tensor{\sigma}_a}}
\; ,
\end{equation}
where $\diffusivity_{Li}$ is the diffusivity of Li ions in the hosting lattice. Note that the transport of dissolved lithium is driven by chemo-mechanical effects, since the lithium flux is proportional to the gradient of Li concentration and to the gradient of hydrostatic pressure.

Guided by Joule effect, a linear law is set as usual for the electrons flow. Accordingly, the current is made proportional to the gradient of the electric potential through the electrical conductivity $\kappa_a > 0$ 

\begin{equation} \label{eq:ohmlaw}
\vect{i}_a = \kappa_a \, \gradient{\phi_a} \, .
\end{equation}

\subsection{Interface conditions} \label{interfaces}

We introduce the boundary values of a generic function $f(\vect{x}, t)$ at the interface between electrodes and electrolyte $I = \partial V_a \cap \partial V_e$ as 
\begin{equation}
\left. f \right|_I^a = \lim_{\vect{x} \in V_a \to I } f(\vect{x}, t) \, , \qquad \left. f \right|_I^e = \lim_{\vect{x} \in V_e \to I } f(\vect{x}, t) \, ,
\end{equation}
and its jump as
\begin{equation}
\label{eq:jump}
 \qquad  \llbracket f \rrbracket = \left. f  \right|_{I}^a - \left. f  \right|_{I}^e \, .
\end{equation}
The electrochemical reaction occurring at the electrodes surface is here modeled through the Butler-Volmer equations \cite{C6CP04142F} without accounting for double layer capacitance as in \cite{DanilovNottenEA2020}. Instead of resolving explicitly the boundary layers between electrodes and electrolyte, the involved phenomena are incorporated in a zero-thickness interface at which the electric potential is discontinuous and the kinetics of the surface electrochemical reaction are defined in terms of the current density $i_{BV}$ as follows
\begin{equation} \label{eq:butler_volmer}
i_{BV} = i_0 \left\{ \exp \left[ \frac{ \alpha_A  \, F \, \eta_S}{R \, T} \right]  - \exp \left[ -  \frac{\alpha_C \, F \, \eta_S }{R \, T}   \right]\right\} \, 
\end{equation}
(see \cite{C5CP03836G} for a comprehensive treatment of the subject).
In the Butler-Volmer equation \eqref{eq:butler_volmer}, $\alpha_A$ and $\alpha_B$ are positive kinetic constants, and $\eta_S$ is the surface overpotential. The exchange current density, $i_0$, is function of the concentration of lithium at the interface, in the form \cite{malaveEA2014b}
\begin{equation} 
\label{eq:exchange_current}
i_0 = K_S \, F \, \left( \left. c_{\text{Li}^+} \right|_I^{e} \right)^{\alpha_A} \, \left( c_{\text{Li}}^{max} - \left. c_{\text{Li}} \right|_I^{a}  \right)^{\alpha_A} \, \left( \left. c_{\text{Li}} \right|_I^{a} \right)^{\alpha_C} \, ,
\end{equation}
while the surface overpotential is defined as
\begin{equation}\label{eq:surface_overpotential}
\eta_S = \llbracket \phi \rrbracket - U_S \, ,
\end{equation}
with $U_S$ denoting the surface open circuit potential, and $\llbracket \phi \rrbracket$ the jump of the electric potential at the interface defined as the potential at the electrode minus the potential in the electrolyte, as in \eqref{eq:jump}. The surface open circuit potential (OCP) $U_S$ is related to the ideal chemical potential, $\mu_{\text{Li}}$, of lithium at the surface of the active material through the following equation \cite{PurkayasthaMcMeekingCM2012},
\begin{equation} \label{eq:surface_open_circ}
F \, U_S (t)  = \tilde{\mu}_{\text{Li}} - \left. \mu_{\text{Li}} (t)  \right|_I^{a}  \, ,
\end{equation}
where $\tilde{\mu}_{\text{Li}}$ is the chemical potential of lithium of a reference electrode. 
The surface mass flux in normal direction at the electrode/electrolyte interface will be denoted with $h_{BV}$. It is related, through the Faraday's law, to the surface current density in the same direction at the same location,
\begin{equation}
i_{BV} = F \, h_{BV} \, .
\end{equation}
There is no intercalation of X$^-$ ions into the active materials. From the mechanical point of view, compatibility and traction continuity are imposed across the interface $I$. In conclusion, the electro-chemo-mechanical interface conditions can be summarized as follow

\begin{subequations}
\begin{gather}
\vect{h}_{\text{Li}} \cdot \vect{n}_a = - \vect{h}_{\text{Li}^+} \cdot \vect{n}_e =  h_{BV}  \, ,\\
\vect{h}_{\text{X}^-} \cdot \vect{n}_a = 0    \, ,\\
\llbracket \phi \rrbracket = U_S -  \eta_S \, \\
\llbracket \vect{u} \rrbracket= \vect{0} \, , \\
 \tensor{\sigma}_a \vect{n}_a   =  - \tensor{\sigma}_e \vect{n}_e    \, .
\end{gather}
\end{subequations}


\section{Electrode design}
\label{sec:pb_def}

Reshaping the fictitious planar battery represented in Fig. \ref{fig:planar_scheme} in order to achieve optimal battery performances is the goal of the electrode design.  
%
%
To this aim, a weak form has been derived after rewriting the governing equations of each battery component in dimensionless form: the detailed calculations can be found in appendix \ref{app:FEM}. The weak form has been discretized in space via the Finite Elements Method (FEM), while the evolution in time has been tracked with the Backward Euler scheme, as in  \cite{SalvadoriEtAlJPS2015,SalvadoriEtAlJPS2015b}. Finally, the resulting nonlinear problem has been implemented in the commercial FEM software ABAQUS, by writing a \textit{User Element Subroutine}. 

\subsection{Material parameters}

\begin{figure}[htbp]
\centering
 \includegraphics[width=13cm]{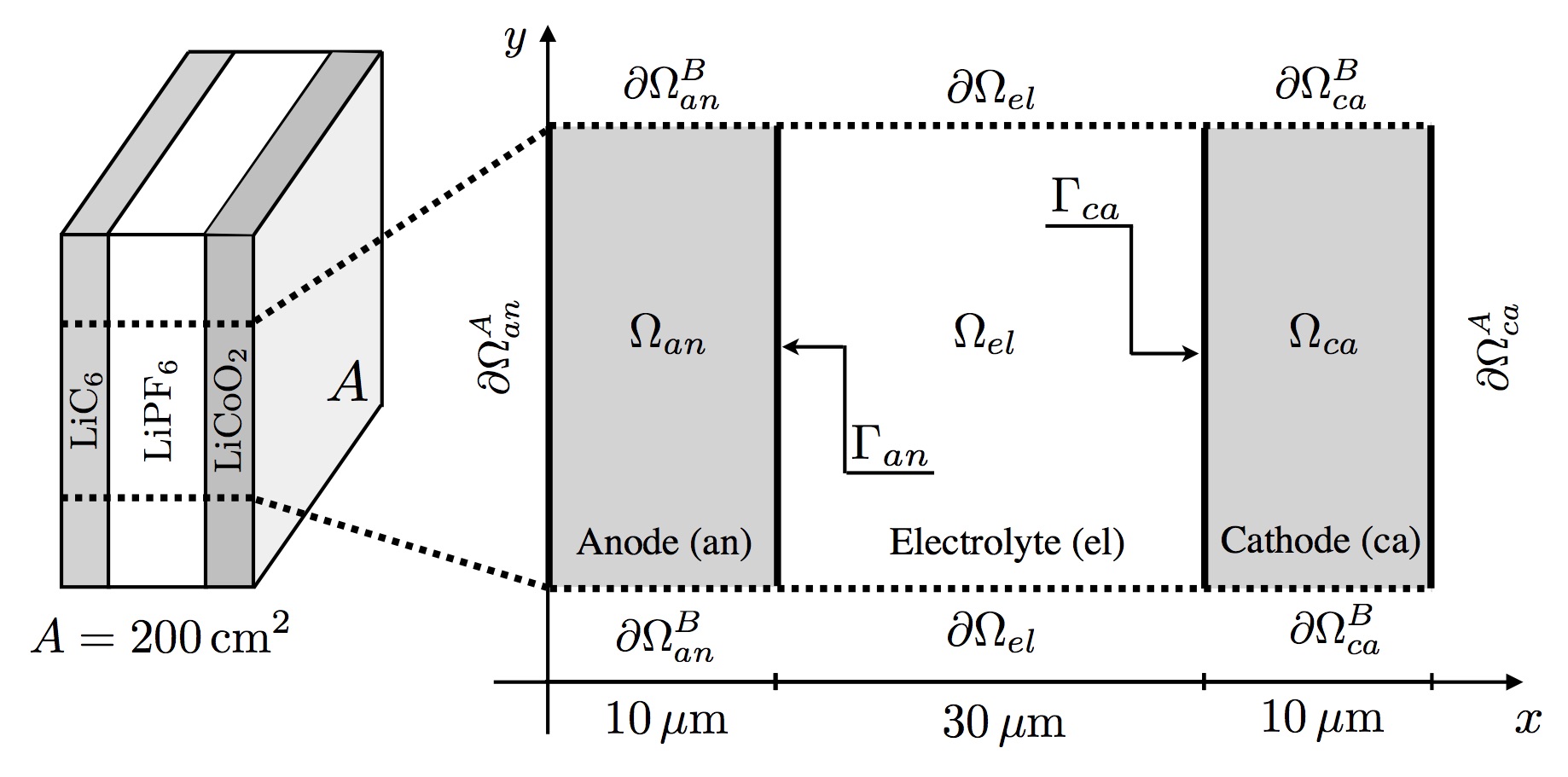}
\caption{\em{The basic configuration (i.e. $n=0$) for the numerically analyzed Li-ion cell.}}
\label{fig:planar_scheme}
\end{figure}

The electrodes initial configuration consists of a 10 $\mu \text{m}$ thick graphite anode, and a 10 $\mu \text{m}$  LiCoO$_2$ cathode separated by a 30 $\mu \text{m}$ separator and LiPF$_6$ liquid electrolyte. The cross sectional area, i.e. the one depicted with $A$ in fig. \ref{fig:planar_scheme} and {\em{not to be confused neither with the area of the interface between electrodes and electrolyte, nor with the electrode net cross area $A_{net}$ that will be defined later on}}, is assumed 200 $\text{cm}^2$ as representative of a commercial cylindrical battery \cite{DanilovNottenEA2008}. The material parameters, taken from the literature, are listed in Table \ref{tab:mat_par}.
Subscripts $_{an}$ and $_{ca}$ will identify LiC$_6$ anode and LiCoO$_2$ cathode from now on. 

\begin{figure}[htbp]
\centering
 \includegraphics[width=10cm]{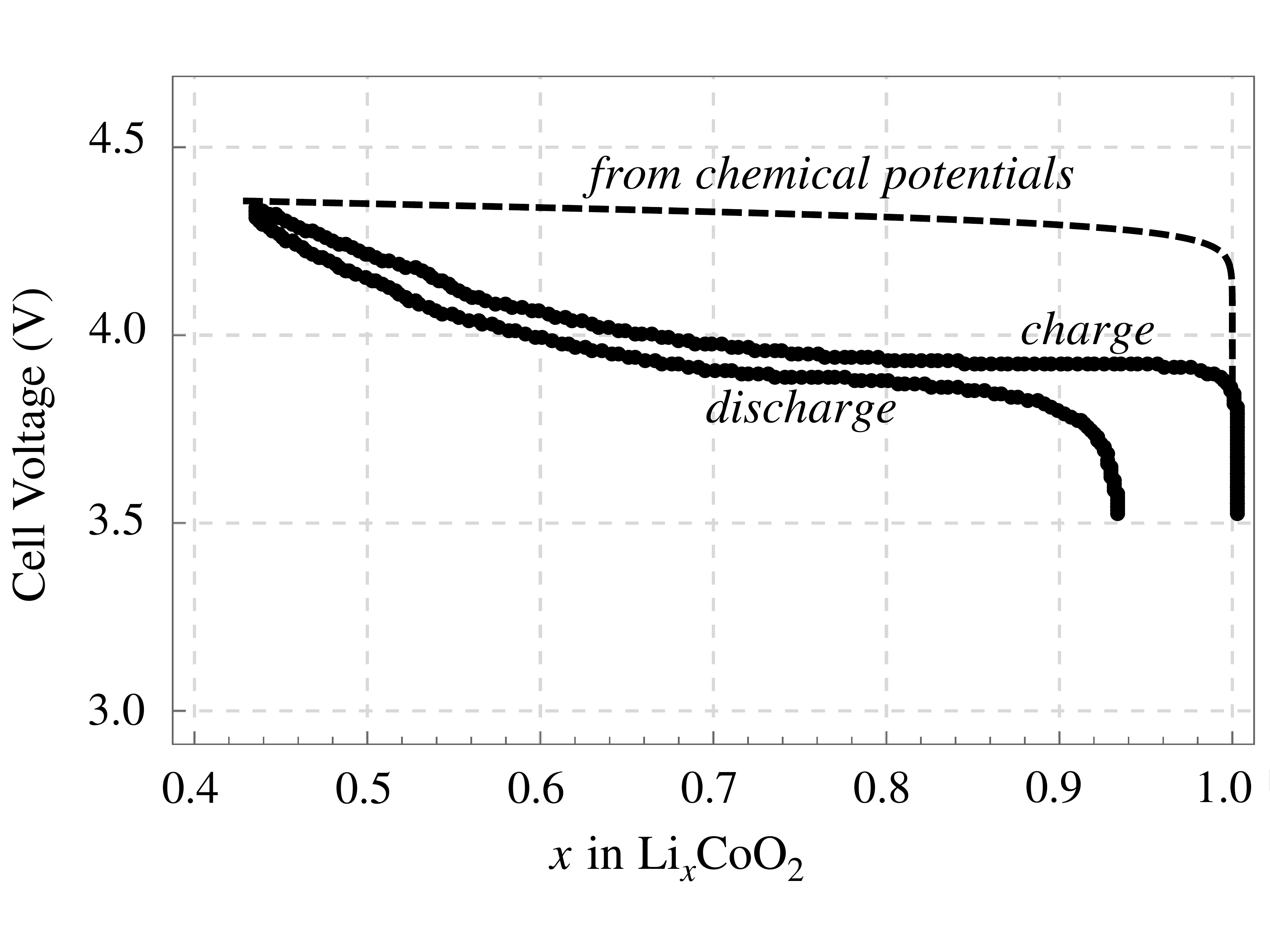}
\caption{\em{Comparison between the theoretical OCP from eq. \eqref{eq:surf_OCP} and the one experimentally measured in \cite{Reimers01081992}. The discrepancies are due to the phase transformation in ${\rm Li}_x{\rm CoO_2}$ that have been highlighted in \cite{Reimers01081992} and cannot be captured by an ideal (purely entropic) model.}}
\label{fig:OCP_comparison}
\end{figure}

Denote with $0<x<1$ the amount of lithium in Li$_x$CoO$_2$. Since in normal battery operations $x$ ranges between $x=0.5$ and $x=1.0$, the theoretical capacity of the battery can be estimated from the volume of the positive electrode $V_{\text{LiCoO}_2}$ as
\begin{equation}
V_{\text{LiCoO}_2} \,\frac{c_\text{Li}^{max}}{2}  \, F = 64 \, \text{mAh} \, ,
\end{equation}
where $ F = 26.801 {\rm A·h/mol} $ is Faraday's constant. 
%
The intercalation reactions of lithium in positive and negative electrodes are written as
\begin{equation} \label{eq:int_react_anode}
\ch{\text{Li}^+ + \text{C}_6 + \text{e}^- <>[ discharge ][ charge ] \text{LiC}_6 }
\end{equation}
and
\begin{equation} \label{eq:int_react_cathode}
\ch{\text{CoO}_2 + \text{Li}^+ + \text{e}^- <>[ charge ][ discharge ] \text{LiCoO}_2 }
\; .
\end{equation}
The surface OCP in formula \eqref{eq:surface_open_circ} is given in terms of the ideal chemical potential \eqref{eq:chem_pot} as 
\begin{equation} \label{eq:surf_OCP}
U_S = - \frac{ \mu_\text{Li} }{F}  = - \frac{ \mu_\text{Li}^0 }{F} - \frac{R \, T }{F} \, \text{ln} \left( \frac{c_\text{Li}}{c_\text{Li}^{max} - c_\text{Li}} \right) + \frac{\omega_\text{Li}}{F} \, \trace{\tensor{\sigma}_a}  \, ,
\end{equation}
having taken a vanishing $\tilde{\mu}_\text{Li}$ as usual. The OCP defined in formula \eqref{eq:surf_OCP} depends upon the concentration of dissolved lithium and on the hydrostatic pressure at the electrode surface. 
The value for $\mu_\text{Li}^0$ has been taken as
\begin{equation}
\frac{\mu^0_\text{Li}}{F}  = - 4.3 \, \text{V} \, 
\end{equation}
in order to match the the measured OCP of graphite and lithium cobalt oxide for Li$_{0.5}$CoO$_2$, assuming a stress free electrode, i.e for $c_\text{Li} = c_{\text{Li}}^{max}/2$ and $\trace{\tensor{\sigma}_a}=0$. A comparison between theoretical and experimental OCP is plotted in Fig. \ref{fig:OCP_comparison}.

%

\renewcommand{\arraystretch}{0.5}
\begin{table}
\centering
\begin{tabular}{c c c c c c}  
\hlineB{3.0}
\\
Material Parameters & & ref. & & ref.   \\
\\
\hlineB{3.0}
\tiny
\\
& \bf{LiC$\mathbf{_6}$ anode} & & \bf{LiCoO$\mathbf{_2}$ cathode} &  \\
\\
$c_{\rm Li}^{max} \, \left[ \rm{mol} / \rm{m}^3 \right] $ & $2.64 \times 10^{4}$  & \cite{Bohn2013} & $2.39 \times 10^{4}$  &  \cite{malaveEA2014b} \\
 \\
 $\diffusivity_\text{Li} \, \left[ \rm{m}^2 / \rm s \right]$ &  $3.90  \times 10^{-14}$  & \cite{Bohn2013} &  $5.387  \times 10^{-15}$ & \cite{malaveEA2014b} \\
  \\
 $\omega_\text{Li} \, \left[ \rm{m}^3 / \rm mol \right] $ & $ 1.642  \times 10^{-6}$  & \cite{Bohn2013} &  $ - 5.300  \times 10^{-7}$ & \cite{Mukhopadhyay2014}\\
  \\
 $ E \, \left[  \text{Gpa}  \right]$ & $ 15 $ & \cite{Bohn2013} & $ 370 $  & \cite{malaveEA2014b} \\
 \\
 $ \nu \, \left[ - \right] $ & $ 0.3  $ &  \cite{Bohn2013}  & $ 0.2  $ & \cite{malaveEA2014b} \\
 \\
 $ \kappa \, \left[ \text{S/m} \right]$ & $ 100 $ & \cite{RenganathanEtAl2010} & $ 10 $ &  \cite{RenganathanEtAl2010} \\
\\
\\
& \bf{LiPF$\mathbf{_6}$ electrolyte} & &  &  \\
\\
$c^{max} \, \left[ \rm{mol} / \rm{m}^3 \right] $ & $1.0 \times 10^{4}$  &  \cite{SalvadoriEtAlJPS2015b}& &   \\
 \\
 $\diffusivity_{\text{Li}^+} \, \left[ \rm{m}^2 / \rm s \right]$ &  $2.0  \times 10^{-11}$  &  \cite{SalvadoriEtAlJPS2015b}& &  \\
  \\
 $\diffusivity_{\text{X}^-} \,  \left[ \rm{m}^2 / \rm s \right] $ & $ 3.0  \times 10^{-11}$  &  \cite{SalvadoriEtAlJPS2015b} &   & \\
  \\
 $ E \, \left[  \text{MPa}  \right]$ &  $450  $ & \cite{ChenJAPS2018} &  & \\
 \\
 $ \nu \, \left[ - \right] $ & $0.499$   &  This study & & \\
 \\
 $ \permittivity_r \left[ -  \right]$ & $ 95 $ &  \cite{SalvadoriEtAlJPS2015b} & &   \\
 \\
 \\
 & \bf{LiC$\mathbf{_6}|$LiPF$\mathbf{_6}$  interface} & & \bf{LiCoO$\mathbf{_2}|$LiPF$\mathbf{_6}$  interface} &  \\
\\
$\alpha_A \, \left[ - \right] $ & $0.5$  & \cite{Guo01102011} & $0.5$  &  \cite{Guo01102011} \\
 \\
 $\alpha_C \, \left[ - \right]$ &  $0.5$  & \cite{Guo01102011} &  $0.5$ & \cite{Guo01102011} \\
  \\
 $K_S \, \left[\text{m}^{2.5} \, \text{mol}^{-0.5} \, \text{s}^{-1} \right] $ & $ 1.764  \times 10^{-11}$  & \cite{Guo01102011} &  $ 6.667  \times 10^{-11}$ & \cite{Guo01102011}\\
  \\
\hlineB{3.0}
\end{tabular} 
\caption{\em{Material parameters used in the numerical simulations. The mechanical parameters are given in terms of Young's modulus $E$ and Poisson ratio $\nu$.}}
\label{tab:mat_par}
\end{table}

\medskip

\subsection{Architected geometries} 
\label{sec:geom_bc}

\begin{figure}[htbp]
\centering
		\includegraphics[width=16cm]{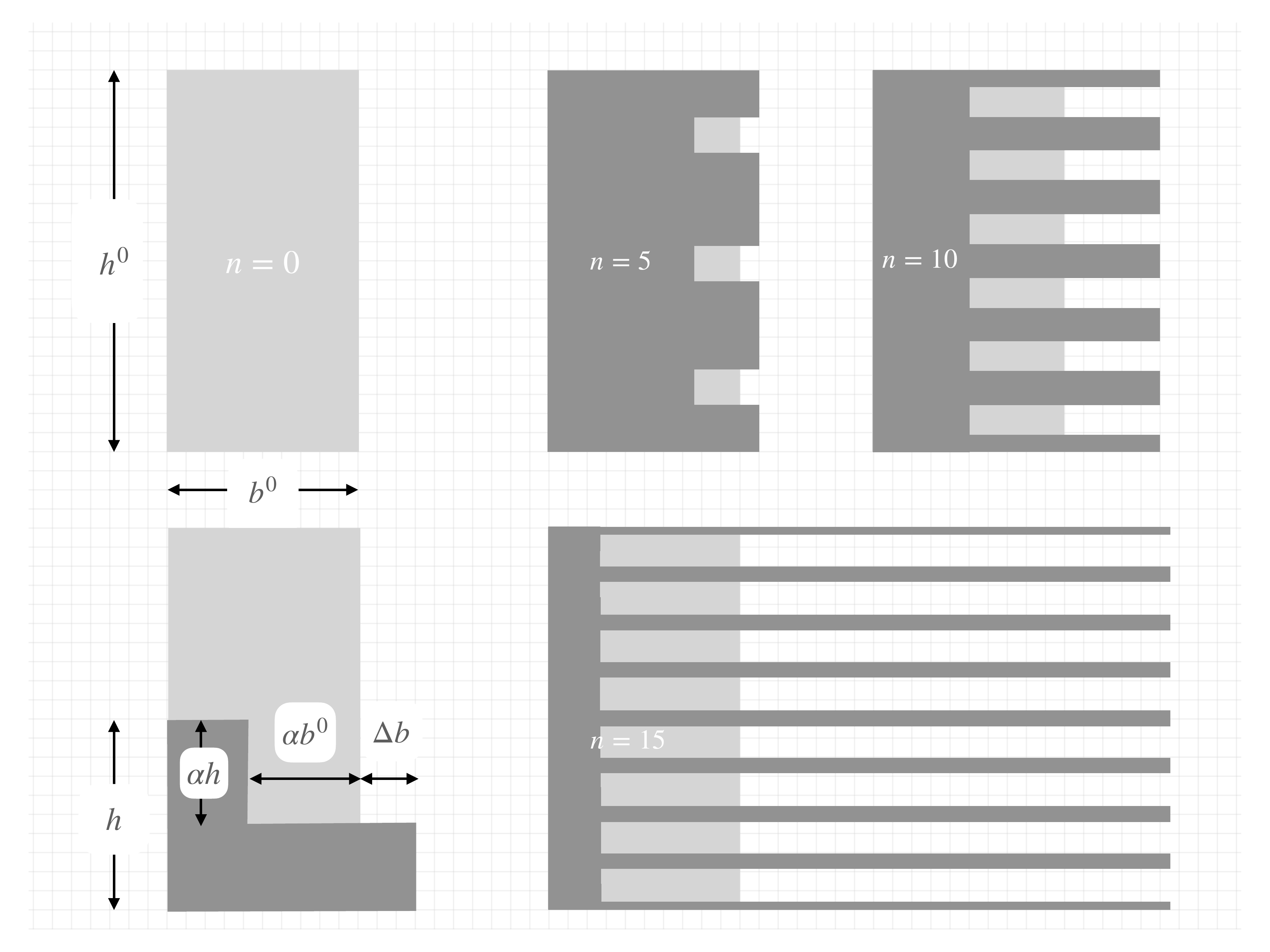}
\caption{\em{The schematic representation of modified electrodes. Light gray shows the planar area at $n=0$.}}
\label{fig:comb_geometry}
\end{figure}

In order to assess the impact of the electrode morphology on the battery performance, either one or both electrodes have been reshaped into the comb-like profile depicted in Fig. \ref{fig:comb_geometry}a.
The {\em{unit cell cross-section}} inherits an index $n$, which controls the degree of optimization. At $n=0$ the selected shape is rectangular, with height $h^0$ and thickness $b^0=10 \rm \mu m$, resembling Fig. \ref{fig:planar_scheme}. The number of unit cells amounts at $n+1$ and the height of each unit cell is 
$$
h(n) = h^0 / ( n+1 ) 
\; ,
$$
so that the total height of the cell remains unchanged and equal to $h^0$.
Figure \ref{fig:comb_geometry}b depicts the shape of the cathode for $n=1, 3, 4, 10$. By increasing the parameter $n$, the protrusions become more pronounced and the comb-like shape more evident. In order to consistently measure the increment of performance with $n$, the capacity and the electrode {\em{net}} area 
kept constant, i.e. 
$$
A_{net} = h^0 \, b^0
\; .
$$ 
The thickness of the separator is fixed at $30 \rm \mu m$ but the thickness of the electrodes change with $n$, though, and a porosity index $\epsilon$ may eventually be associated to $n$ as the ratio of the area occupied by the electrolyte and the net area of the electrode.

\begin{table}[ht]
\begin{center}
\ra{1.3}
\begin{tabular}{l l c r r r c r r c r}
\toprule
$n$ & $\alpha$ && $\Delta b$ & $b_{btm}$ & $b_{top}$ && $h$ & $\alpha h$ && $\epsilon$ \\
\midrule
0 & 0 && 0 & 10  & 10 && 20 & 0 && 0 \\
1 & 0.05 && 0.03 & 10.03 & 9.5 &&  10 & 0.5 && 0.003 \\
5 & 0.25 && 0.83 & 10.83 & 7.5 && 3.33 & 0.833 && 0.077 \\
10 & 0.5 && 5 & 15 & 5 && 1.82 & 0.91 && 0.333 \\
15 & 0.75 && 22.5 & 32.5 & 2.5 && 1.25 & 0.9375 && 0.692 \\
\bottomrule
\end{tabular} 
\caption{Geometric data of the unit cell.}
\label{table:comb}
\end{center}
\end{table}

\begin{figure}[htbp]
\centering
    \begin{subfigure}[!htb]{.3\textwidth}
        \centering
        \includegraphics[width=\textwidth]{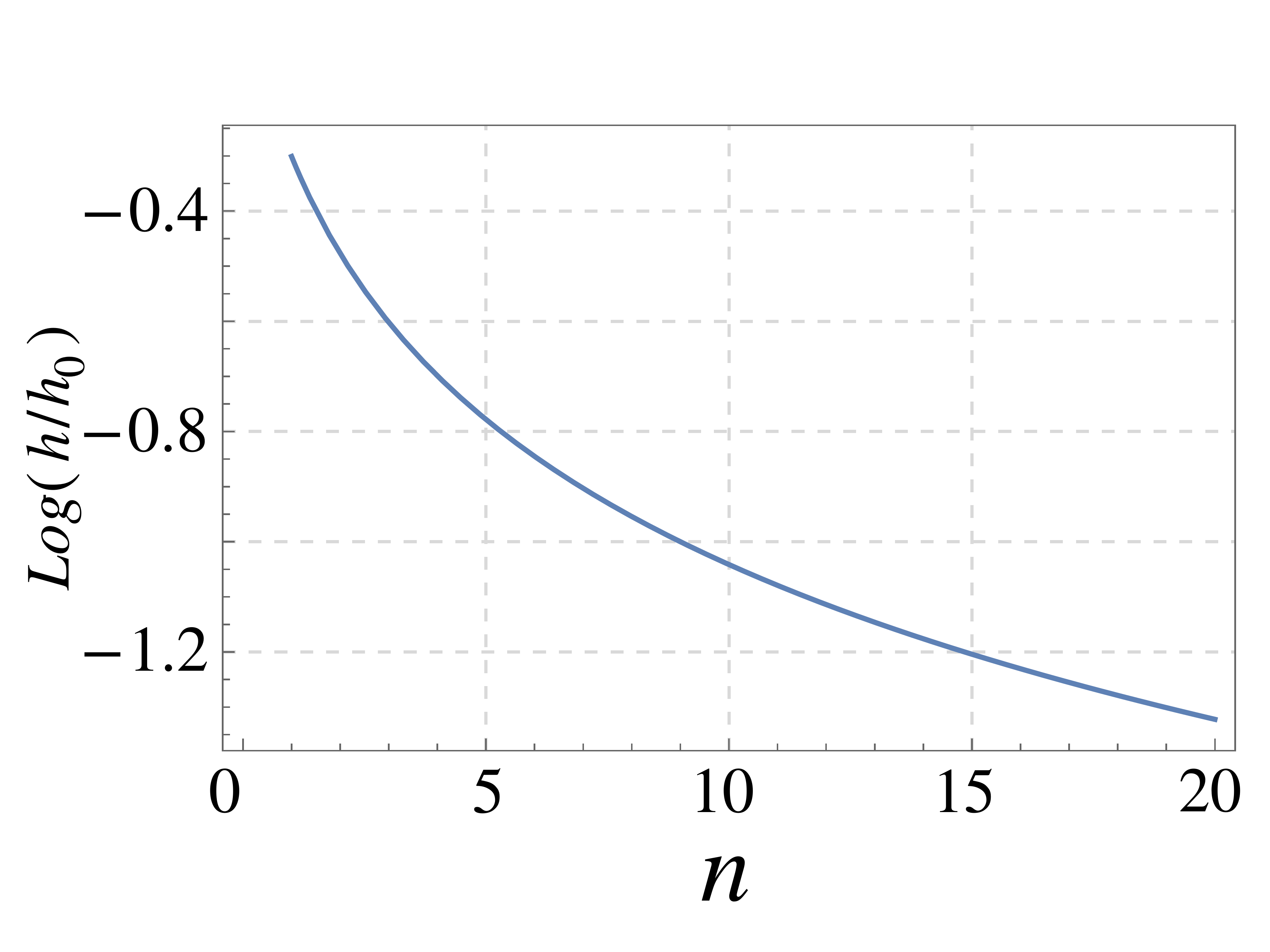}
        \caption{ }
    \end{subfigure} 
    \begin{subfigure}[!htb]{.3\textwidth}
        \centering
        \includegraphics[width=\textwidth]{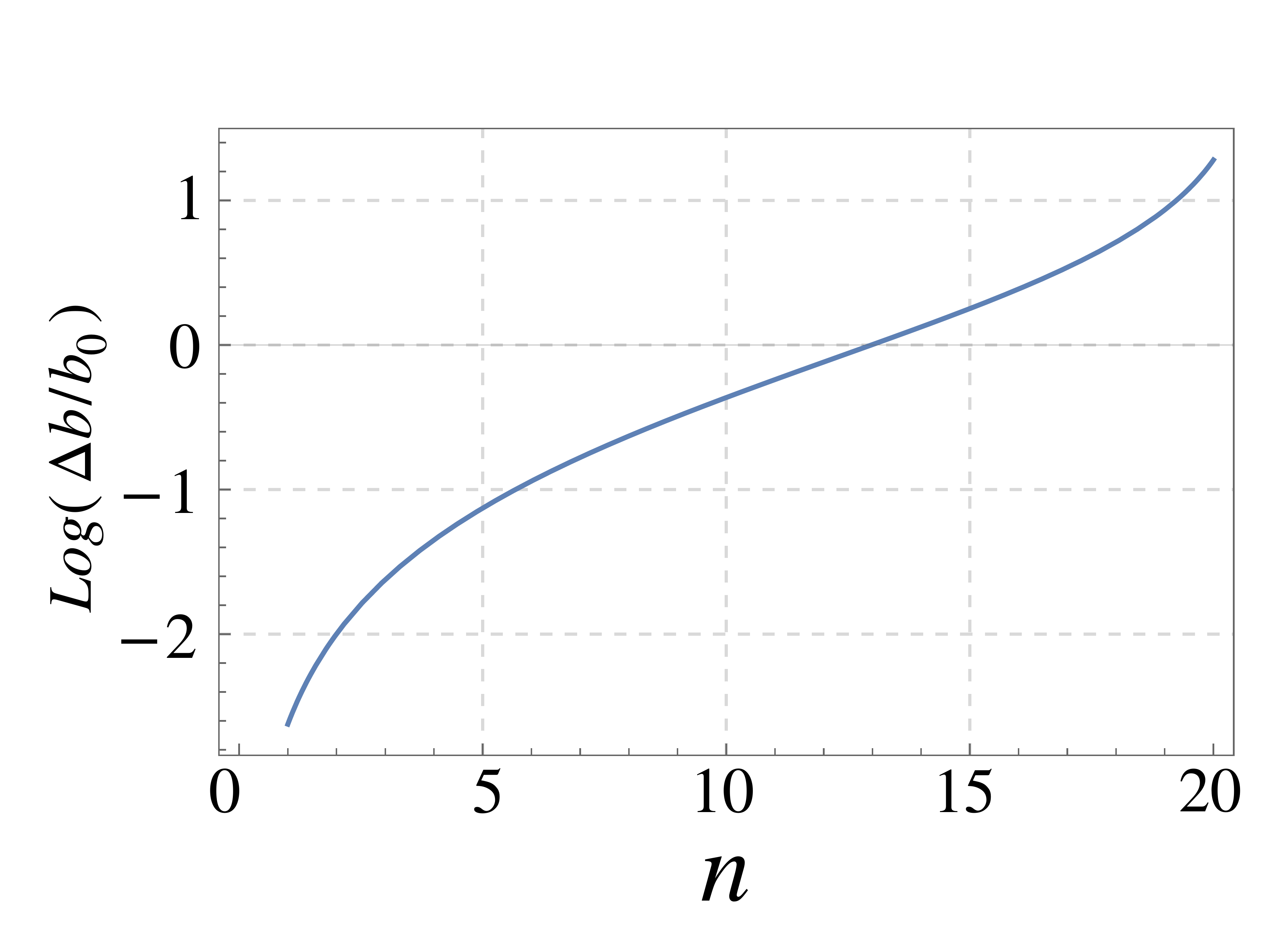}
        \caption{ }
    \end{subfigure} 
    \begin{subfigure}[!htb]{.3\textwidth}
        \centering
        \includegraphics[width=\textwidth]{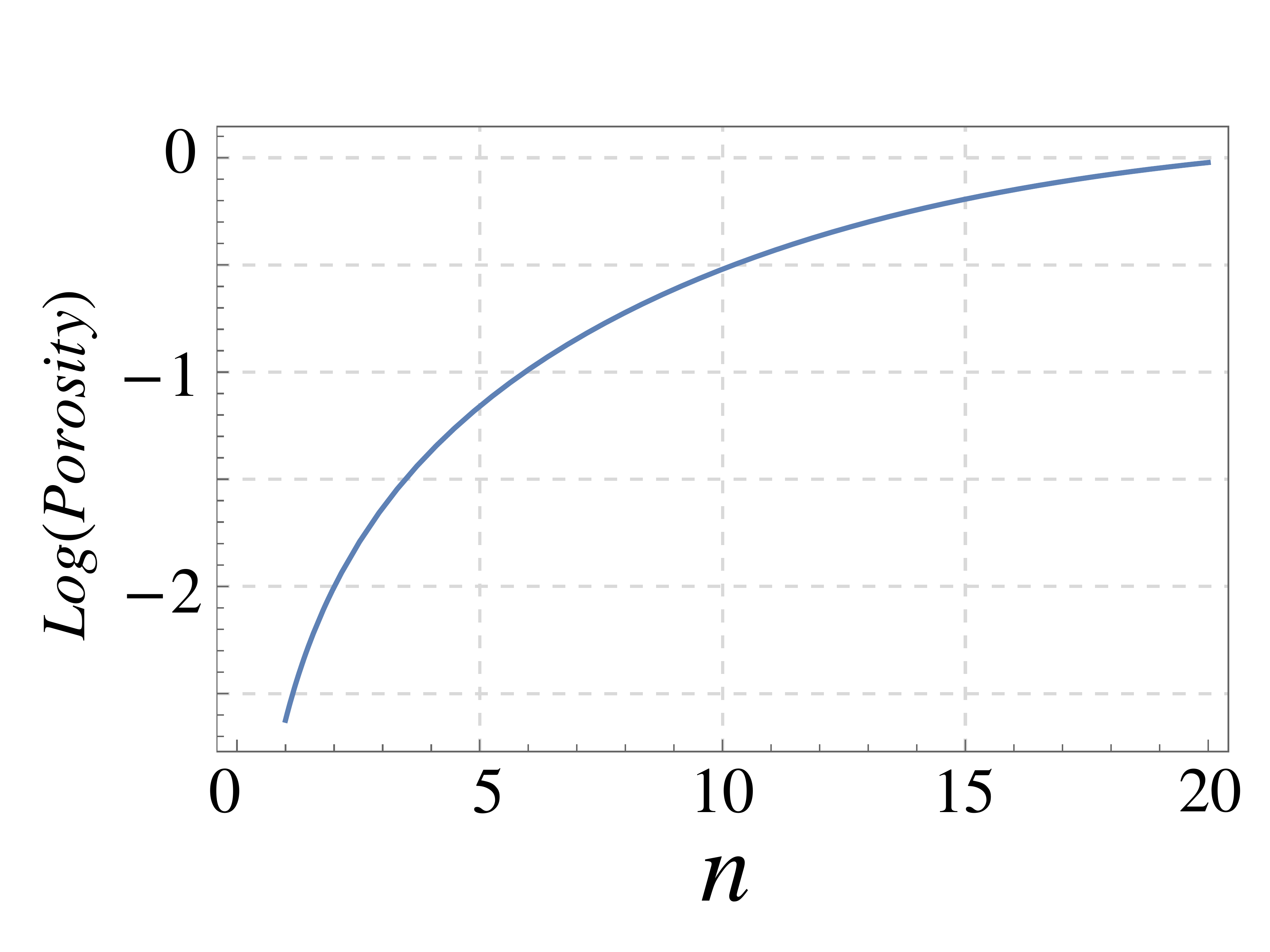}
        \caption{ }
    \end{subfigure} 
\caption{\em{Log plots for the dimensionless ratio $h/h^0$, $\Delta b / b^0$, and for the porosity for the evolution of the electrode geometry with n=0,1,... and $\alpha(n)$ as in (\ref{eq:alpha_ele}). }}
\label{fig:n_comparison}
\end{figure}

The shape of the unit cell cross-section, which evolves from rectangular to an L-shape, is managed by a factor $0 \le \alpha(n) < 1$ and by the amount 
$$
\Delta b(n) = b^0 \; \frac{ \alpha^2 }{ 1 - \alpha } 
\; ,
$$
as depicted in Fig. \ref{fig:comb_geometry}. 
The selection of function $\alpha(n)$ is arbitrary, provided that $\alpha(0) = 0$.
The size of the bottom side of the unit cell, see again Fig. \ref{fig:comb_geometry}, is 
$$
b_{btm}(n) = b^0 + \Delta b  = b^0 \, \left( 1 +  \frac{ \alpha^2 }{ 1 - \alpha } \right)
\; ,
$$ 
whereas the top side length amounts at 
$$
b_{top}(n) = b^0 ( 1 - \alpha(n) ) 
\; .
$$
The total area of the unit cell amounts at 
\begin{equation}
    A^{uc}_{tot}(n) = h \, b_{btm} = A_{net} \; \frac{1}{n+1} \, \left( 1 +  \frac{ \alpha^2 }{ 1 - \alpha } \right)
\end{equation}
hence the total area of the electrode yields
\begin{equation}
    A_{tot}(n) = (n+1 ) \, A^{uc}_{tot}(n) = A_{net} \; \left( 1 +  \frac{ \alpha^2 }{ 1 - \alpha } \right)
\end{equation}
and finally the porosity ratio comes out
\begin{equation}
    \epsilon(n) = 1 - \frac{A_{net}}{A_{tot}(n)} =  \frac{ \alpha^2 }{ 1 - \alpha + \alpha^2 }
    \; .
\end{equation}
The new comb-like geometry adds porosity to electrodes and improves the intercalation by two mechanisms, by increasing the surface area and by narrowing the thickness of the electrodes in the combs, so to increase the amount of lithium that can be effectively stored. The second feature turns out to be particularly relevant at high C-rates.
As for function $\alpha(n)$,
we chose 
\begin{equation}
\label{eq:alpha_ele}
    \alpha(n) = \frac{n}{n_{max}}
\end{equation}
in our simulations, with $n_{max} = 20$ the (arbitrary chosen) upper bound of the number of combs. 
The geometric data for a few values of $n$ are reported in Table \ref{table:comb}.
Logarithmic plots for the dimensionless ratio $h/h^0$, $\Delta b / b^0$, and for the porosity after such a choice are depicted in Fig. \ref{fig:n_comparison}.



%

\subsection{Boundary and initial conditions} 

For the sake of clarity, the electrodes boundary is divided into two parts, $\partial \Omega^A$ and $\partial \Omega^B$, as depicted in Fig. \ref{fig:planar_scheme}. The boundary conditions are imposed on the boundary of both the electrodes, $\partial \Omega_{an}$ and $\partial \Omega_{ca}$, and electrolyte $\partial \Omega_e$. Making reference to Fig. \ref{fig:planar_scheme}, all conditions on $\left\{ \partial \Omega_{el} \cup  \partial  \Omega^B_{an} \cup  \partial \Omega^B_{ca} \right\} $ will be vanishing, in this way enforcing a periodicity in the $y$ direction {\em{of Taylor type}}.


The battery response will be evaluated upon constant current discharging at different C-rates. To this end, a uniform discharging current is prescribed on both sides of the battery as follows
\begin{equation}
 \vect{i}_{an} \cdot \vect{n}_{an}  = -\frac{ I (t)}{A}  \qquad  \text{on} \;  \partial \Omega^A_{an} \, , \qquad \vect{i}_{ca} \cdot \vect{n}_{ca}  =  \frac{ I (t)}{A}  \qquad  \text{on} \;  \partial \Omega^A_{ca} \,  ,
\end{equation}
where $A$ is the battery cross-sectional area (see Fig. \ref{fig:planar_scheme}) and $\vect{i}$ refers to the (ionic) current flowing through the battery. In order to make initial and boundary conditions compatible with thermodynamic equilibrium at $t=0$, $I$ is tuned in time as
\begin{equation}
\label{eq:current}
I(t) =\left( 1- \text{e}^{-t} \right) \, I_{1.0} \, C_r \, ,
\end{equation}
where $I_{1.0} = 3.2 \, \text{A} / \text{m}^2$ is the electric current corresponding to 1 C-rate discharging, while $C_r$ is the discharge rate. The flux of current is zero on $\left\{ \partial \Omega_{el} \cup  \partial  \Omega^B_{an} \cup  \partial \Omega^B_{ca} \right\} $
\begin{align*}
 \vect{i}_{an} \cdot \vect{n}_{an}  = 0  \qquad  \text{on} \;  \partial \Omega^B_{an} \, , \qquad  \vect{i}_{ca} \cdot \vect{n}_{ca}  = 0  \qquad  \text{on} \;  \partial \Omega^B_{ca} \, , \qquad  \vect{i}_{el} \cdot \vect{n}_{el}  = 0  \qquad  \text{on} \;  \partial \Omega_{el} \, .
 \end{align*}
In line with battery morphology and processes, ionic species Li$^+$ and X$^-$ cannot flow through the external boundary  so that 
\begin{subequations}
\begin{gather*}
 \vect{h}_\text{Li} \cdot \vect{n}_{an}  = 0  \qquad  \text{on} \;  \partial \Omega_{an} \, , \qquad  \vect{h}_\text{Li} \cdot \vect{n}_{ca}  = 0  \qquad  \text{on} \;  \partial \Omega_{ca} \, ,
 \\
   \vect{h}_{\text{Li}^+} \cdot \vect{n}_{el}  = \vect{h}_{\text{X}^-} \cdot \vect{n}_{el} = 0  \qquad  \text{on} \;  \partial \Omega_{el} \, .
 \end{gather*}
\end{subequations}
Assuming the battery case to be infinitely rigid, the cell expansion/contraction is prevented during battery operations. Thus the following mechanical boundary conditions have been set: 
\begin{subequations}
\begin{align*}
u_x = 0  \qquad  \text{on} \; \left\{ \partial \Omega^A_{an} \cup  \partial \Omega^A_{ca} \right\} \, , \qquad u_y = 0  \qquad  \text{on} \; \left\{ \partial \Omega_{el} \cup  \partial  \Omega^B_{an} \cup  \partial \Omega^B_{ca} \right\} \, .
 \end{align*} 
\end{subequations}

Initial conditions have been imposed for species concentration in both cathode and electrolyte at $t=0$. When the battery is first assembled, all the Lithium resides in the positive electrode, the anode being Lithium free when the battery is fully discharged. Thus the battery has to be charged first, so that Li ions can be transferred from the positive to the negative electrode. 
Due to the electrochemical properties of LiCoO$_2$, only half of its theoretical capacity is usually displaced to the anode and used in battery operations \cite{Reimers01081992}. Therefore, the initial concentration of Li ions in the electrodes is

\begin{equation*}
\Big. c_\text{Li} \Big|_{t=0}  = 1.2 \times 10^{4} \text{mol} / \text{m}^3   \qquad  \text{in} \;  \Omega_{an} \cup \Omega_{ca} \, ,  \\
\end{equation*}

\noindent
since the electrodes have the same volume. The initial concentration of ionic species in the electrolyte is taken from \cite{SalvadoriEtAlJPS2015b}, i.e.

\begin{equation*}
\Big. c_{\text{Li}^+} \Big|_{t=0}  = \Big. c_{\text{X}^+} \Big|_{t=0} = 1.5 \times 10^{3} \text{mol} / \text{m}^3 \qquad  \text{in} \;   
\Omega_{el} \, .
\end{equation*}

\subsection{Numerical methodology}

As already pointed out, the weak form \eqref{eq:weak_form} enclosed in appendix \ref{app:FEM} is discretized in space and time prior to the implementation in the commercial software ABAQUS as a user element subroutine. In this way, the resulting non-linear problem is solved through the Newton-Raphson algorithm in a monolithic scheme by ABAQUS inner algorithmic machinery. 

\begin{figure}[htbp]
\centering
 \includegraphics[width=17cm]{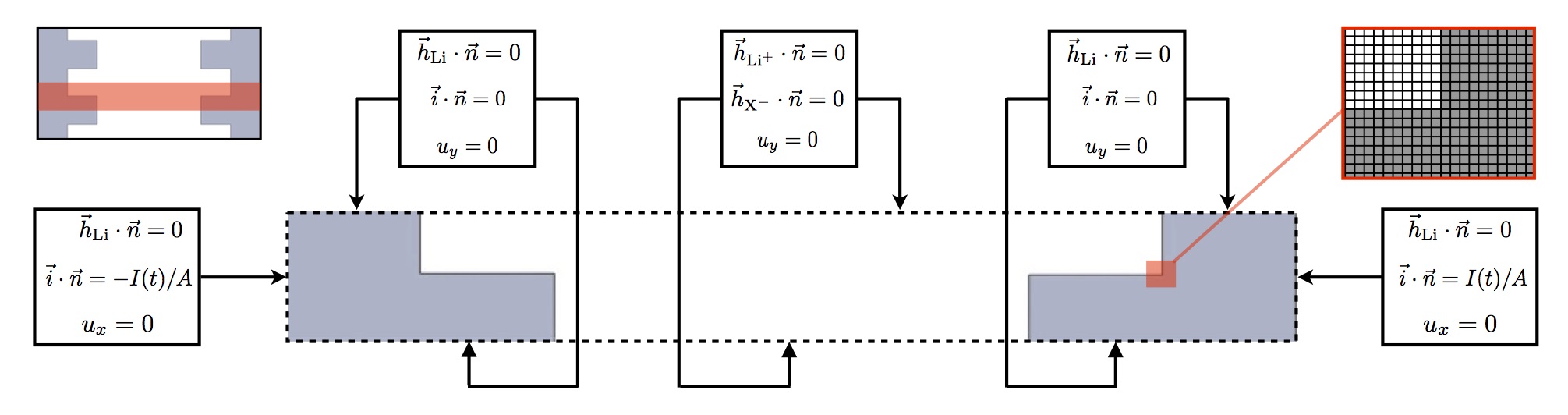}
\caption{\em{Computational domain adopted for the numerical simulations along with a schematic representation of the boundary conditions used.}}
\label{fig:FEM_scheme}
\end{figure}

The numerical implementation of the conditions at the interface between active materials and electrolyte are of particular interest in these electro-chemo-mechanical systems. In this work we implemented zero-thickness elements between electrodes and electrolyte. For this sake, we have extended the numerical approach used for cohesive mechanics (see \cite{GeersCohesive2008,ORTIZPANDOLFI,MULTIZONE,PARK2012239} among others) to the electro-chemo-mechanical interfaces at hand using conformal FE discretization of active materials and electrolyte at the interfaces, i.e. the nodes at interface must superpose (see Fig. \ref{fig:FEM_scheme}). 

The computational domain, which can be reduced to the unit cell reported in Fig. \ref{fig:FEM_scheme} for reasons of symmetry and periodicity, has been discretized using 4-node bilinear elements. Due to the regular morphology of the battery components, a structured FE grid has been generated with the mesh-generator Gmsh \cite{gmshreference}.  

\begin{figure}[htbp]
\centering
 \includegraphics[width=10cm]{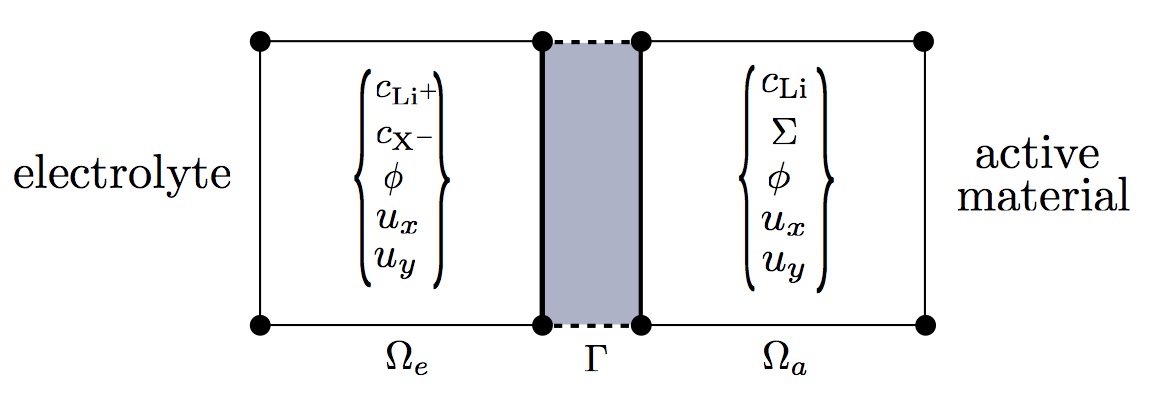}
\caption{\em{Unknown fields in the electrolyte and in the electrodes. Since the nodes of the interface elements belong either to the electrolyte or to the electrodes, the nodal unknowns in the interface elements follow from the ones implemented in the bulk elements.}}
\label{fig:int_elem}
\end{figure}

Three distinct User Elements have been developed for the implementation in ABAQUS: one for the active materials, one for the electrolyte, and one for the interfaces. All these elements have five degrees of freedom as depicted in Fig. \ref{fig:int_elem}. In particular, the nodal unknowns are $\{ c_{\text{Li}^+}$, $c_{\text{X}^-}$, $\phi_e$, $u_x$, $u_y \}$ in the electrolyte and $\{ c_{\text{Li}}$, $\Sigma$, $\phi_e$, $u_x$, $u_y \}$ in the electrodes. Since the nodes of the interface elements belong either to the electrolyte or to the electrodes, the nodal unknowns in the interface elements follow from the ones implemented in the bulk elements.   

\section{Outcomes and discussion}

In all simulations we considered a current-discharge regime \eqref{eq:current} for $\rm C-rates = 1, 2, 4, 8$. The response of the battery with planar electrodes, see Fig. \ref{fig:planar_scheme}, is discussed as the one-dimensional reference configuration $n=0$.  

\subsection{Electro-chemo-mechanical response of the planar configuration} 
\label{sec:planar_battery}

\begin{figure}[htbp]
\centering
    \begin{subfigure}[!htb]{.49\textwidth}
        \centering
        \includegraphics[width=\textwidth]{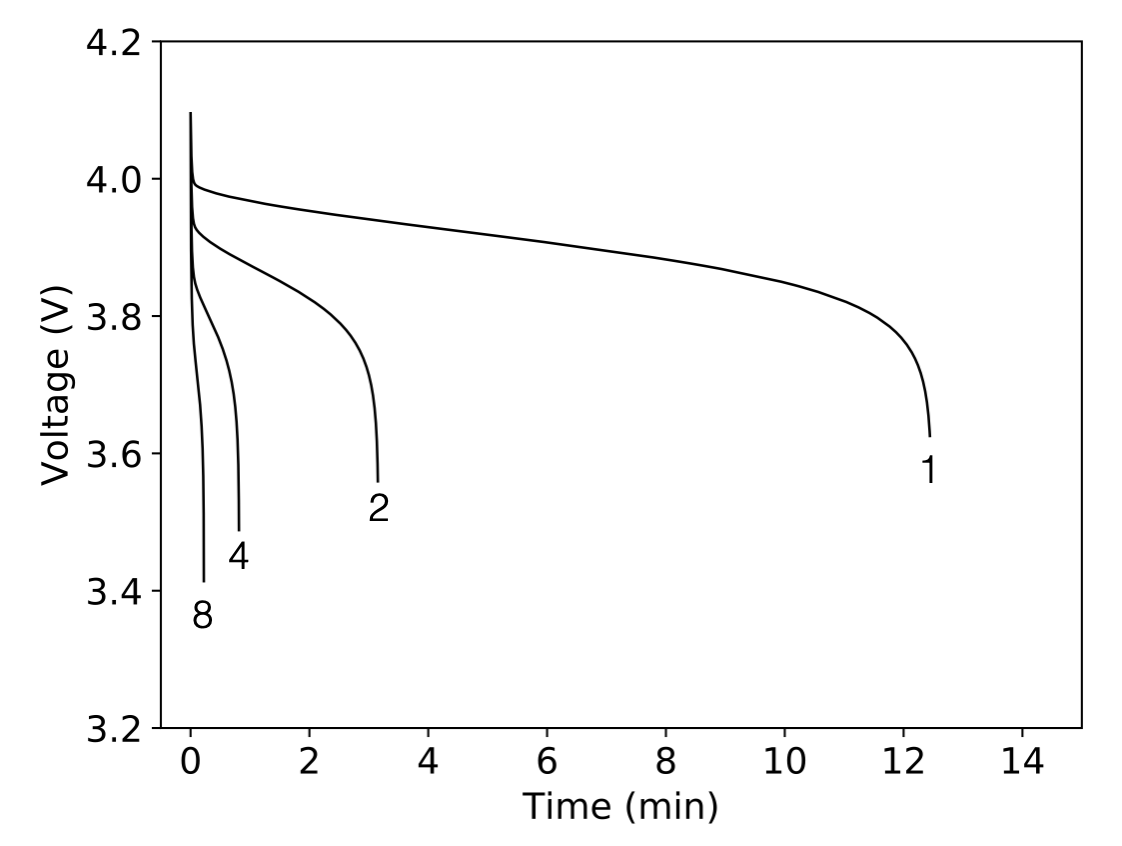}
        \caption{Voltage vs time }
    \end{subfigure}\hfill%
    \begin{subfigure}[!htb]{.49\textwidth}
        \centering
        \includegraphics[width=\textwidth]{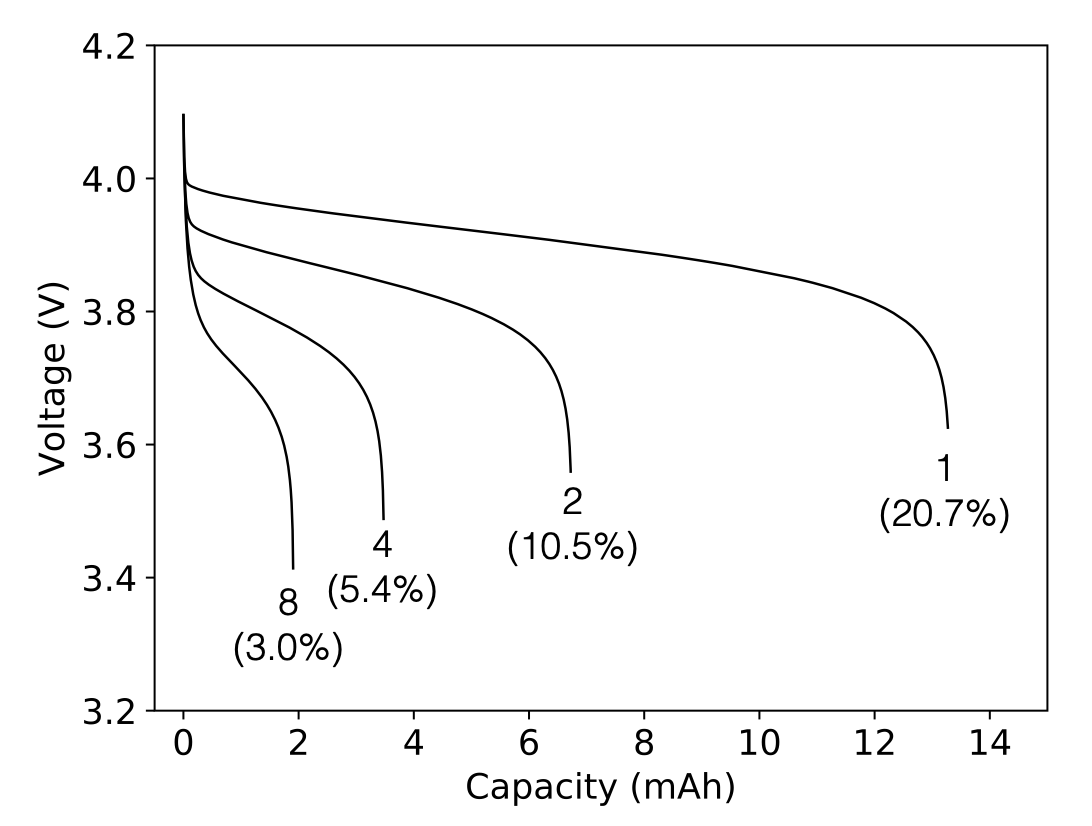}
        \caption{Voltage vs capacity }
    \end{subfigure}
    %
%
%
\caption{\em{Simulated voltage profiles of the planar battery during constant-current-discharging at C-rates = 1, 2, 4, and 8. Fig. (b) reports in brackets the ratio of the extracted charge(capacity) at the end of the discharge over the theoretical capacity of the battery, i.e. 64 mAh. }}
\label{fig:voltage_n=0}
\end{figure}

\textit{The battery voltage} - 
Figure \ref{fig:voltage_n=0}a shows the predicted voltage profiles as a function of time, while in Fig. \ref{fig:voltage_n=0}b the battery voltage is plotted against the extracted charge, the integral in time of the current that flows across the battery cell. Independently upon the discharging rate, the voltage profiles show three distinct phases: at first the voltage drops sharply (IR drop), then the discharge proceeds showing a plateau, finally the process ends with another sudden voltage drop. It is worth recalling that here we do not make use of the experimental OCP, as carried out in many numerical works, rather we relate the surface OCP in the Butler-Volmer equation to the chemical potential of Li ions in the electrodes, as shown in Eq. \eqref{eq:surface_open_circ}. 

 As it emerges from experimental data \cite{Danilovetal2011,DanilovNottenEA2020}, the battery voltage is highly dependent on the C-rates. Indeed the operating voltage attains lower values for higher discharging rate as a result of higher battery overpotentials. The amount of extracted charge is dependent on the discharge rate as well (see Fig.  \ref{fig:voltage_n=0}b): the higher the discharge rate, the lower the extracted charge. For quantitative comparisons, the ratio between the simulated extracted charge and the theoretical capacity is reported in brackets in Fig.  \ref{fig:voltage_n=0}b. This percentage can be regarded as a measure of the efficiency of the battery.  For the planar configuration, the battery performs poorly (for instance, at 8 C-rate, only $3 \%$ of the theoretical capacity is used; even for 1 C-rate the amount of extracted charge is only about $20 \%$ of the theoretical capacity).

\medskip
 
\begin{figure}[htbp]
\centering
    \begin{subfigure}[!htb]{.49\textwidth}
        \centering
        \includegraphics[width=\textwidth]{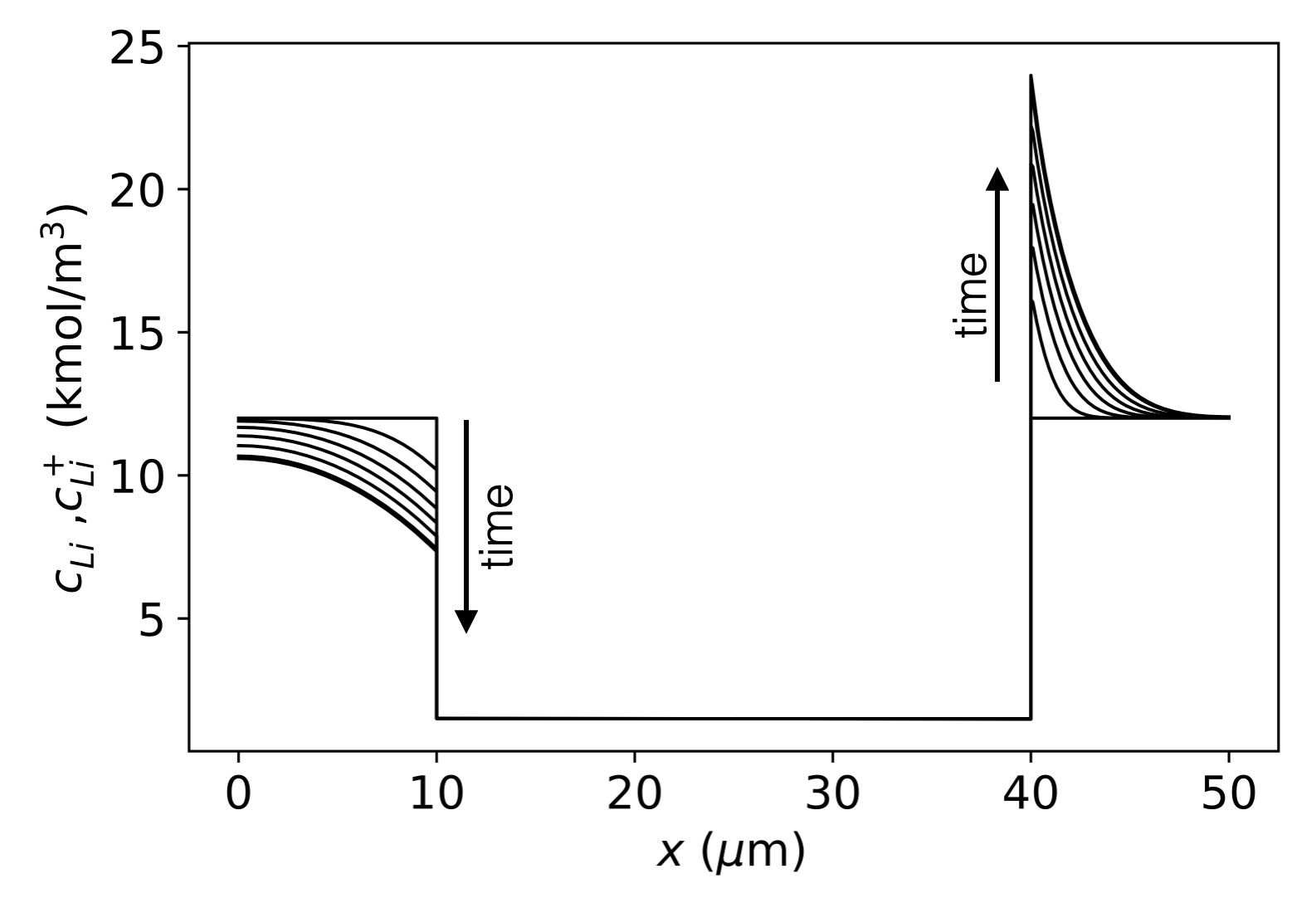}
        \caption{ }
    \end{subfigure} 
    \begin{subfigure}[!htb]{.49\textwidth}
        \centering
        \includegraphics[width=\textwidth]{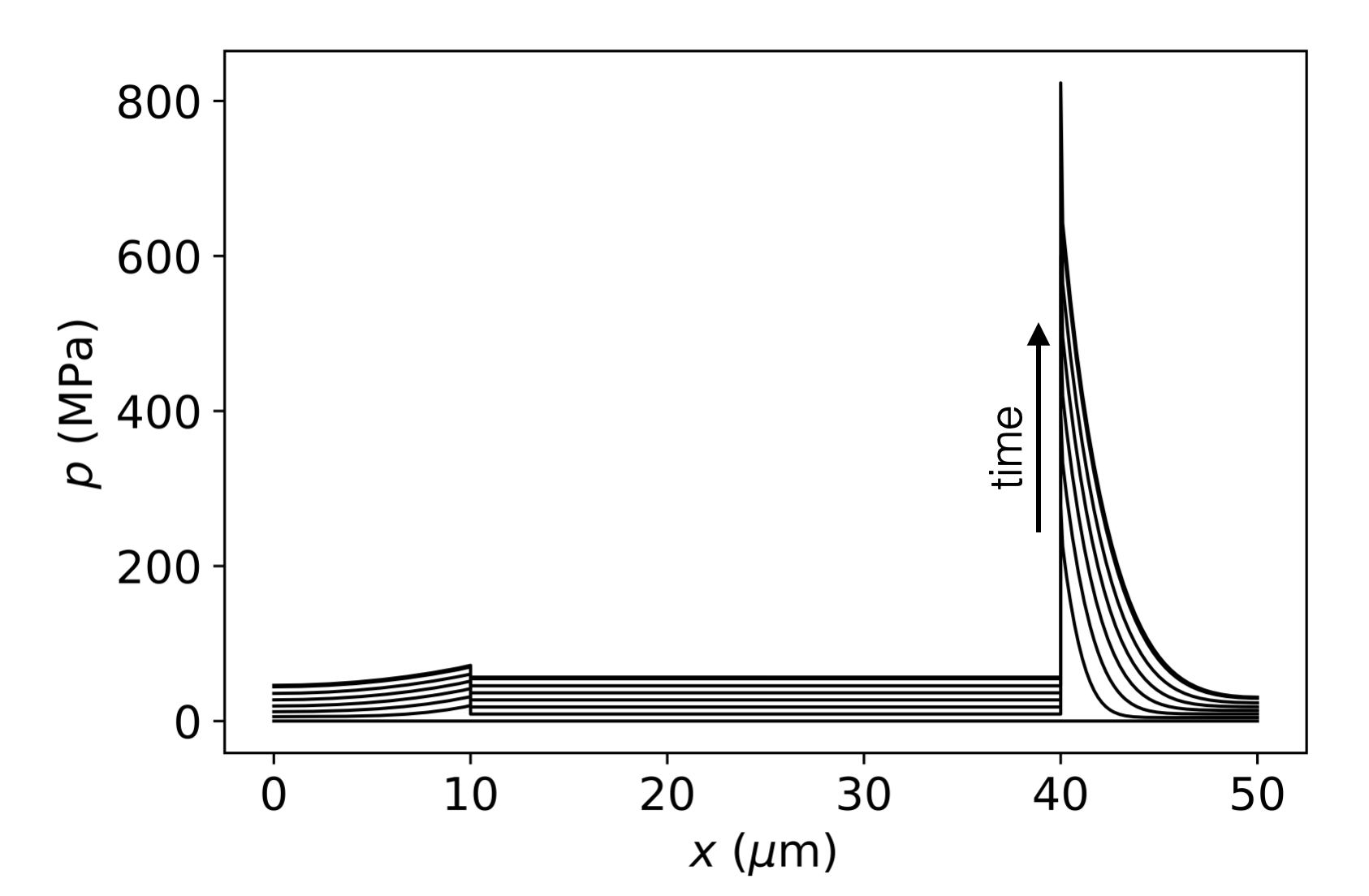}
        \caption{ }
    \end{subfigure} 
%
\caption{\em{(a) Lithium concentration  and (b) hydrostatic pressure in the planar battery as a function of x coordinate at time t=0, 2, 4, 6, 8, 12 minutes, and at the end of the process for 1.0 C-rate discharging.}}
\label{fig:chemo_mechanical_n=0}
\end{figure}

\textit{The chemo-mechanical response} - Figure \ref{fig:chemo_mechanical_n=0}a plots the evolution in time of Li ions concentrations within the planar battery at 1.0 C-rate discharging. During discharging, the concentration of lithium progressively increases in the cathode, accumulating near the electrolyte interface, while the opposite occurs at the negative electrode. The simulation ends after 12.5 minutes, when lithium in the cathode reaches its saturation limit in correspondence of $\Gamma_{ca}$ ($c_{\text{Li}}^{max} = 2.39 \times 10^{4} \, \text{mol}/\text{m}^3$). Indeed, further current flow, in the prescribed regime, is prevented by saturation of lithium in the positive electrode in eq. \eqref{eq:exchange_current}. The latter is thus the limiting factor for the performance of this planar Li-ion battery. Owing to the much higher ionic diffusivity, the concentration profile of Li ions is essentially flat in the electrolyte. The overall amount of Li ions dissolved in the electrolyte is conserved since lithium is consumed at $\Gamma_{ca}$ with the same velocity as it is generated at $\Gamma_{an}$, and the charge is conserved.

\begin{SCfigure} [] [htb!]
\centering
\includegraphics[width=7.5cm]{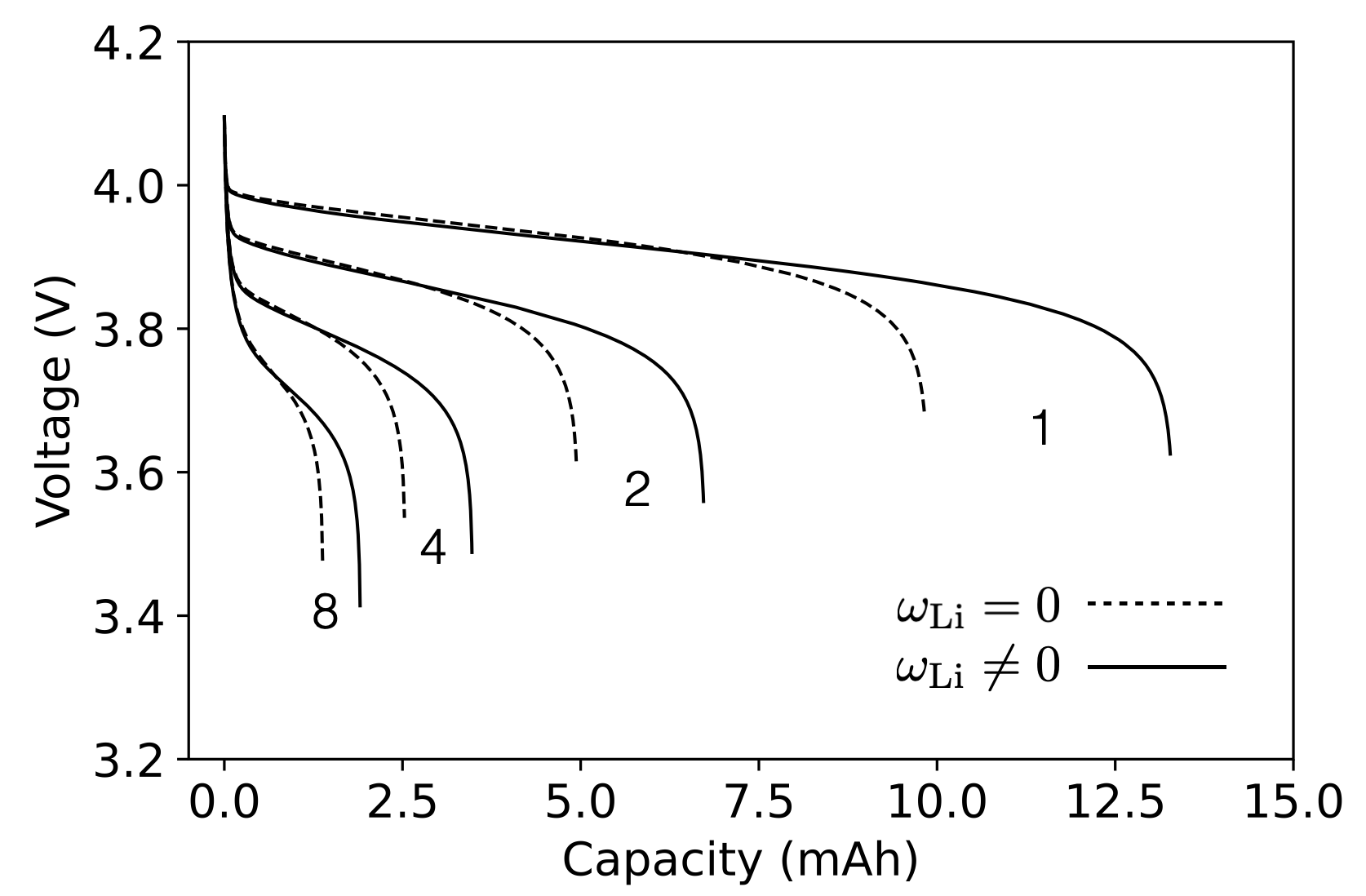}
\caption{\em{Comparison between the simulated voltages in the planar battery obtained though fully coupled chemo-mechanical analysis ($\omega_\text{Li} \neq 0$), and by neglecting the mechanical effects ($\omega_\text{Li} = 0$). The battery has been simulated upon constant-current-discharing at 1.0, 2.0, 4.0, 8.0 C-rates.}}
\label{fig:voltage_mech}
\end{SCfigure}

Figure \ref{fig:chemo_mechanical_n=0}b depicts the simulated mechanical response in terms of hydrostatic pressure, i.e. $p = \frac{1}{3} \text{tr} \left[ \tensor{\sigma} \right]$. Since no external forces are prescribed on the battery case, the stress evolution is triggered by the volumetric strain induced by lithium intercalation and de-intercalation in the electrodes. The separator acts passively, contrasting the volume change in the electrodes with its mechanical stiffness. As shown in Fig. \ref{fig:chemo_mechanical_n=0}b, the pressure, null at $t=0$, increases with time in each component of the battery. On one hand, $p$ is essentially uniform in the separator with magnitude smaller that $100$ Mpa. On the other, the stress distribution is not uniform in the electrodes and attains its maximum value at interface $\Gamma_{ca}$. The stress evolution of $p$ in the electrodes merely emanates from the distribution of lithium, as dictated by equation \eqref{eq:stress_def}. 
The pressure is positive, i.e. tensile, in both positive and negative electrodes because of the value of parameters $\omega_\text{Li}$. In fact, experimental evidence show \cite{lyu2021overview} that LiC$_6$ shrinks upon de-lithiation and LiCoO$_2$ shrinks with lithiation. 
The impact of the stress evolution in the voltage response of the battery is highlighted in Figure \ref{fig:voltage_mech}. The voltage profiles of Fig. \ref{fig:voltage_n=0}b are compared with the ones obtained by neglecting the mechanical effects, i.e. prescribing $\omega_\text{Li} = 0$ in both electrodes. The numerical evidence show that the chemo-mechanical coupling influences the duration of the discharging process, and consequently, the amount of extracted charge at the end of the process. This relies on the definition of the chemical potential in Eq. \eqref{eq:chem_pot}, from which the stress state influences the transport of lithium and the surface OCP in the Butler-Volmer equation (see Eq.s \eqref{eq:Li_flux} - \eqref{eq:surf_OCP}). These outcomes show that the mechanics of intercalating electrodes has a major influence on the operating voltage of Li-ion batteries and cannot be neglected.

\subsection{Electrodes morphologies} \label{sec:cathode_morph} 

In section \ref{sec:arch_cathode}, the positive electrode LiCoO$_{2}$ is reshaped into the comb-like morphology detailed in section \ref{sec:geom_bc}. This morphological change impacts on the amount of surface of the cathode in contact with the electrolyte, i.e. the electrochemical charge transfer reaction takes place on a greater surface compared to the planar battery simulated in Section \ref{sec:planar_battery}. However, as stated in Section \ref{sec:geom_bc}, the new morphology does not alter neither the volume of active material nor the theoretical capacity of the battery. 

We proceeded in section \ref{sec:arch_both} by modifying the geometry of both anode and cathode concurrently: in each battery configuration, the morphology of both electrodes is mirrored. Akin to section \ref{sec:arch_cathode}, the battery geometry is defined by index $n$, without altering neither the volume of active material nor the theoretical capacity of the battery.

\medskip

\subsubsection{Architecting the cathode}
\label{sec:arch_cathode}

\begin{figure}[htbp]
\centering
    \begin{subfigure}[!htb]{.49\textwidth}
        \centering
        \includegraphics[width=\textwidth]{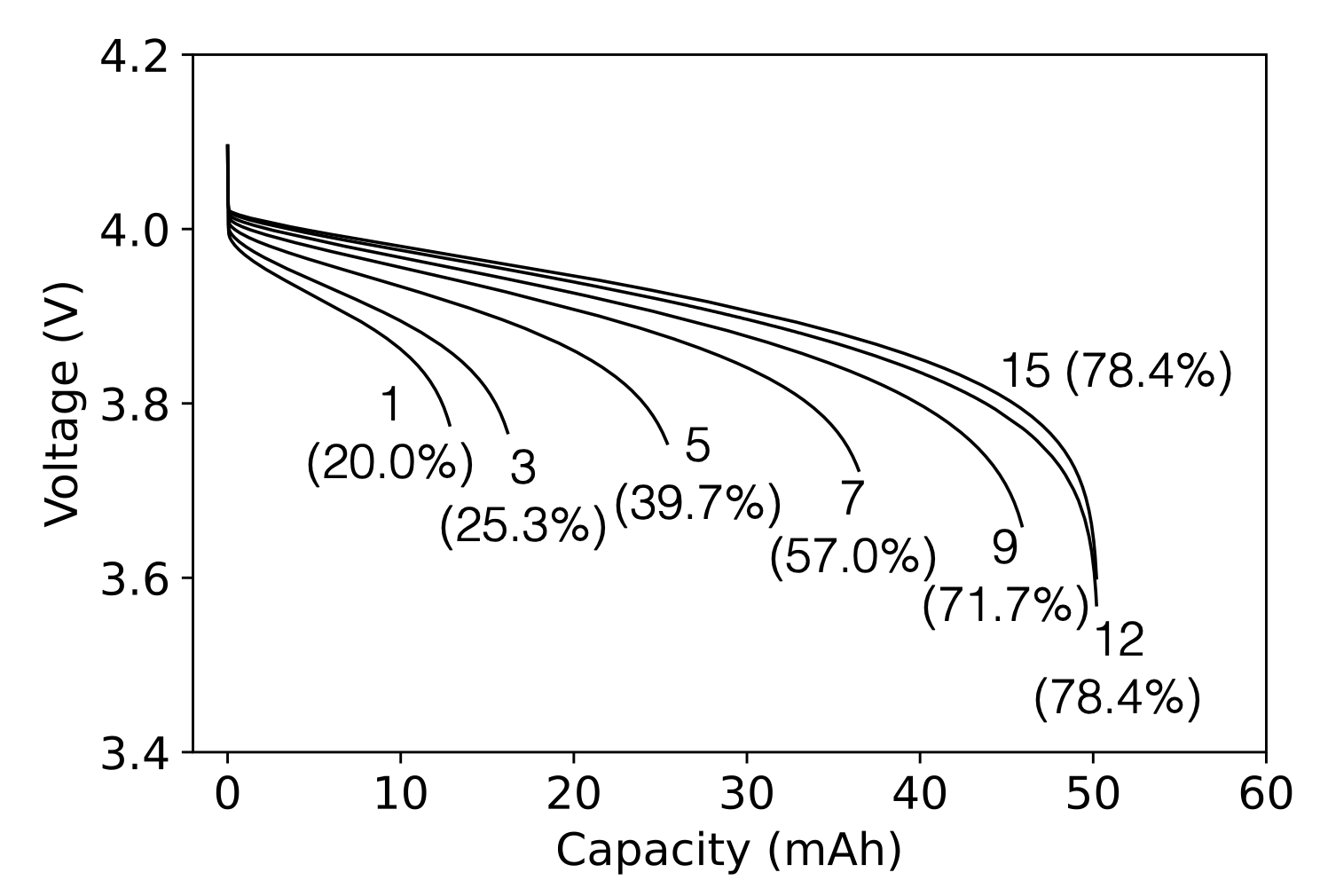}
        \caption{ }
    \end{subfigure} 
    \begin{subfigure}[!htb]{.49\textwidth}
        \centering
        \includegraphics[width=\textwidth]{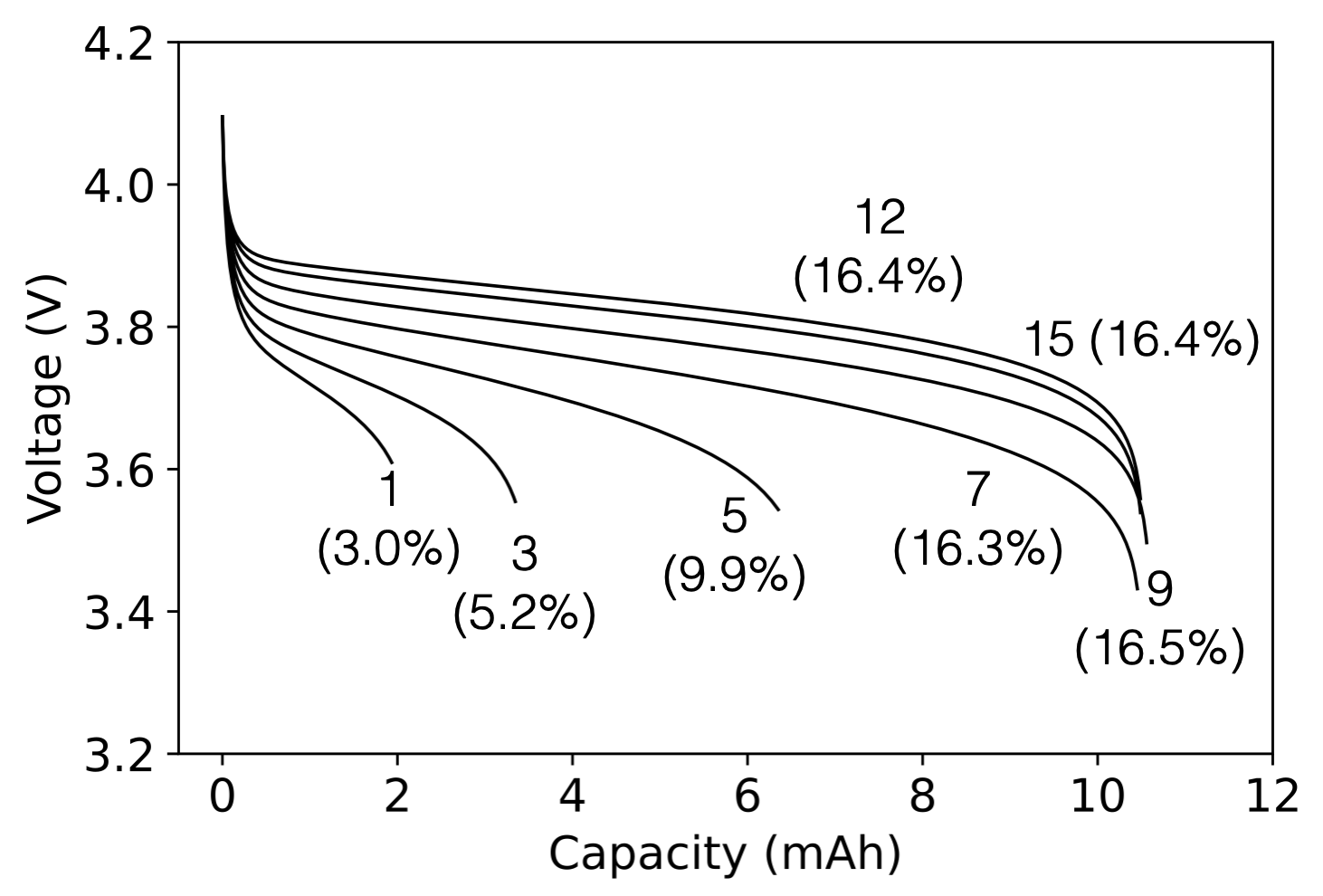}
        \caption{ }
    \end{subfigure} 
%
\caption{\em{Influence of cathode morphology on the simulated voltage for 1.0 (a) and 8.0 (b) C-rate discharging. Each voltage profile is labelled by the integer n, i.e. the constant that identifies the shape of cathode (see Fig. \ref{fig:comb_geometry}). The battery efficiency, reported in parenthesis, is computed as the ratio between the extracted charge at the end of the discharging and the theoretical capacity. }}
\label{fig:voltage1}
\end{figure}

\textit{The battery voltage} - 
Figure \ref{fig:voltage1} depicts the simulated voltage of the battery for different cathode morphologies, with $n$ ranging from 1 to 15. The discharge current \eqref{eq:current} is the same for each analysis. For both C- rates 1 and 8, the higher the number of combs the higher the operating voltage of the battery. Accordingly, the amount of extracted charge is influenced by the electrodes morphology. The rise in the battery efficiency can be seen clearly by comparison with the reference planar configuration. At a unit C-rate, the efficiency gradually yet constantly increases up to $n=9$. Afterwards, for $n\ge10$, the capacity of the battery remains basically steady at 76.56\%. Similar outcomes are observed for C-rate 8 under the same operating conditions, although the extracted charge is much lower.

\medskip

\begin{figure}[!htbp]
\centering
\includegraphics[width=16.5cm]{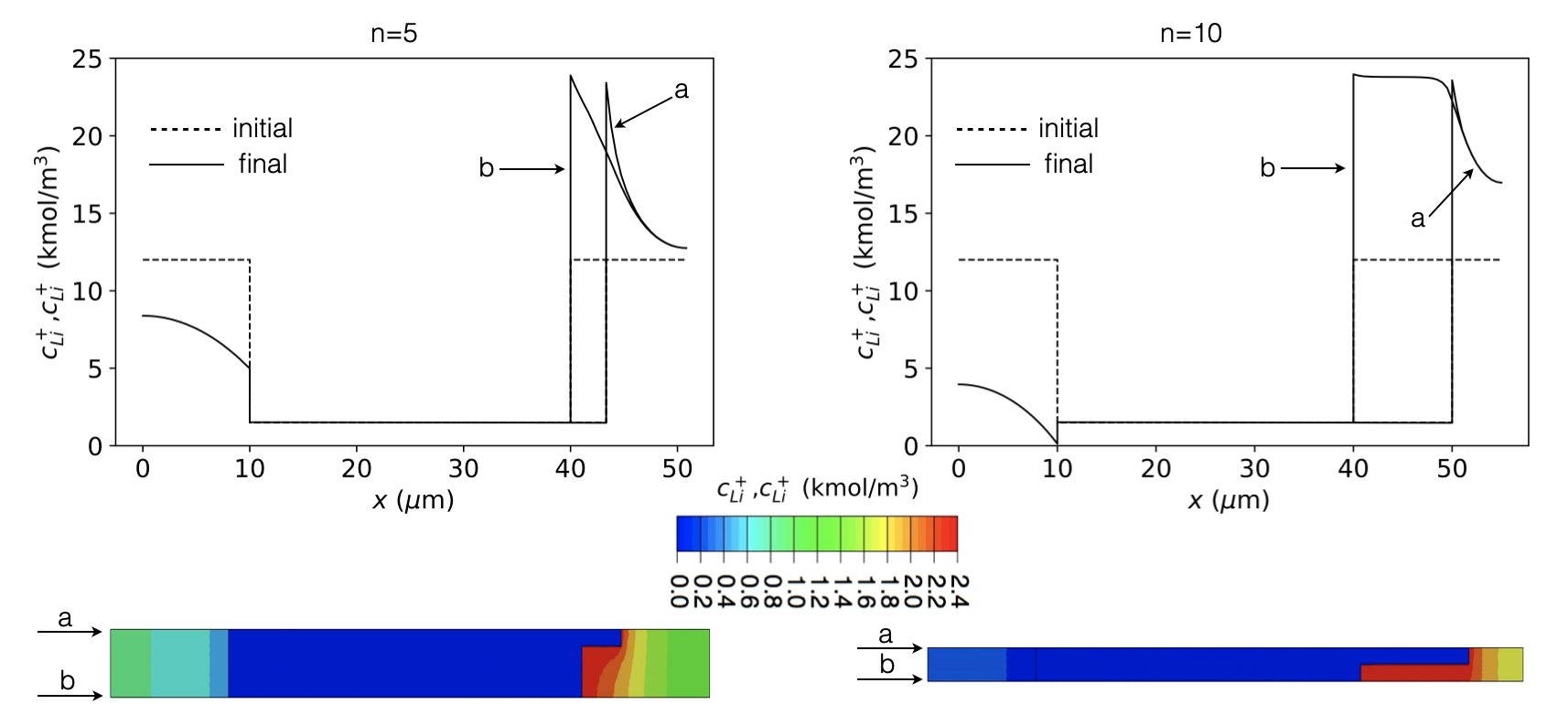}
\caption{\em{Comparison between the lithium distribution in the battery at the end of 1.0 C-rate discharging for cathode morphology $n=5$ and $n=10$.}}
\label{fig:cLi1}
\end{figure}

\textit{The chemo-mechanical response} - The effect of the cathode morphology on lithium concentration is evaluated at 1C-rate for $n=1$ as well as $n=10$ and depicted in Fig. \ref{fig:cLi1}. As expected, the higher the index $n$ the more uniform the lithium distributes in the positive electrode. For $n=1$, 
the lithium intercalated in the cathode accumulates at the electrolyte interface, while the 
lithium in the anode shares with the reference battery ($n=0$) the same profile. The battery discharge ends due to the saturation limit on the cathode interface $\Gamma_{ca}$. Hence, as for the planar electrode configuration, the cathode is the limiting factor for the battery cell on discharge. 

The circumstances change at $n>10$. As shown in Fig.  \ref{fig:cLi1}b, lithium does not saturate in the positive electrode anymore. Rather, the LiC$_6$ becomes the new limiting factor, because the depletion of lithium at the anode at $\Gamma_{an}$ is induced, preventing any further flow of current. This evidence explains why increasing the porosity of the cathode by index $n$ does not provide any further improvement to the performance, which could be obtained only after reshaping both electrodes. 

\subsubsection{Reshaping both electrodes}
\label{sec:arch_both}


\medskip

\begin{figure}[htbp]
\centering
    \begin{subfigure}[!htb]{.49\textwidth}
        \centering
        \includegraphics[width=\textwidth]{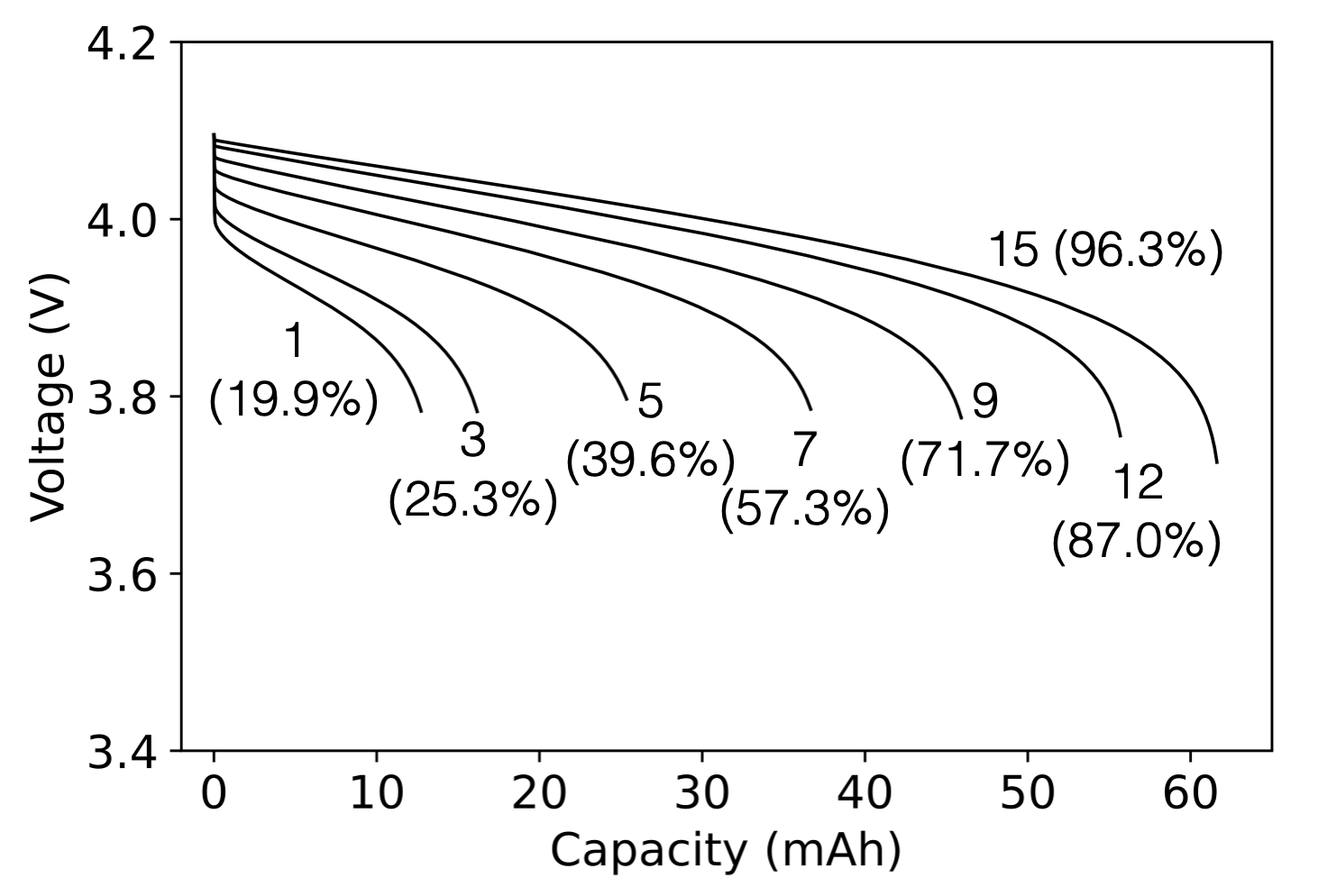}
        \caption{ }
    \end{subfigure} 
    \begin{subfigure}[!htb]{.49\textwidth}
        \centering
        \includegraphics[width=\textwidth]{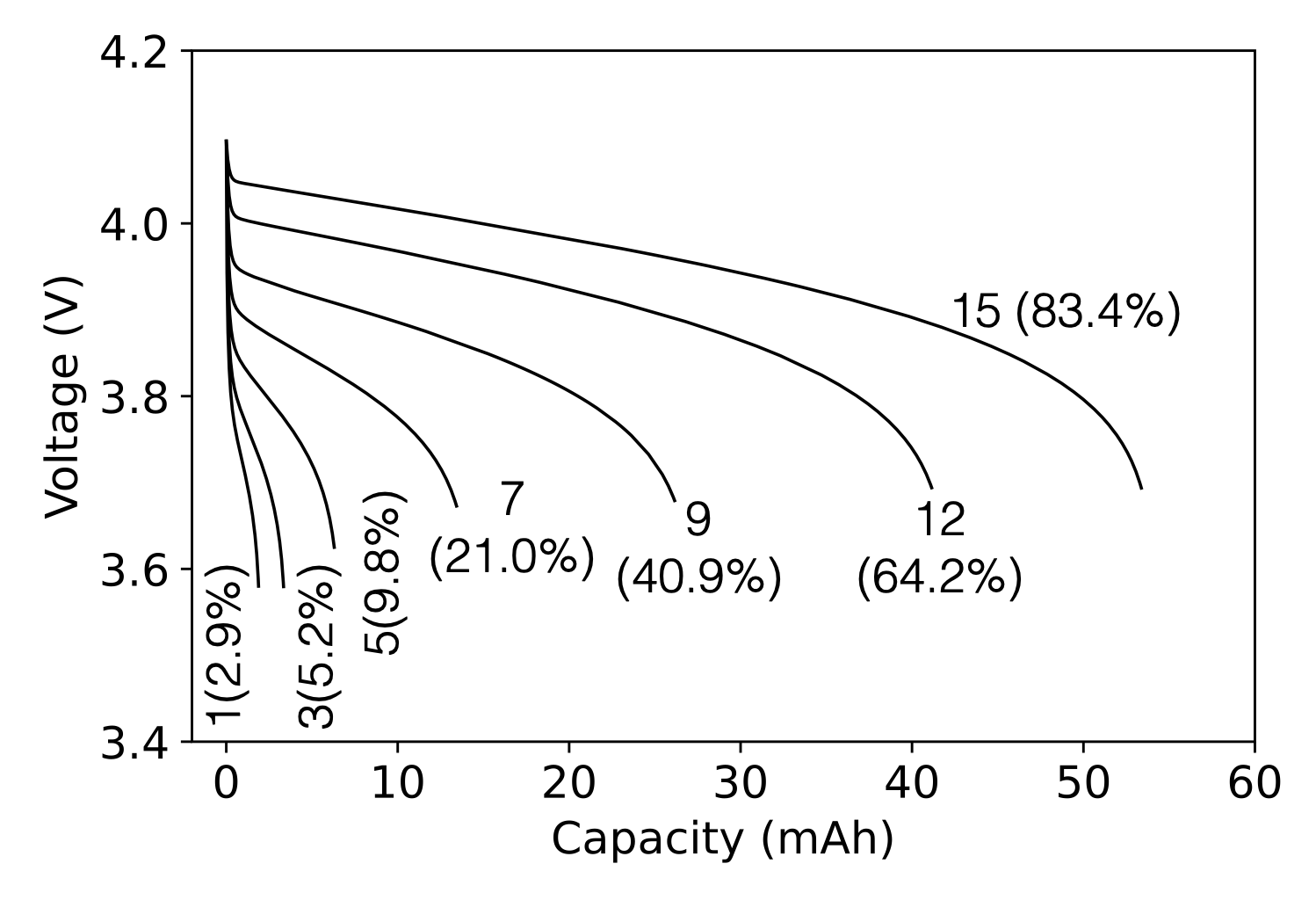}
        \caption{ }
    \end{subfigure} 
%
\caption{\em{Impact of battery morphology on the simulated voltage for 1.0 (a) and 8.0 (b) C-rate discharging. Each voltage profile is labeled by the integer n, i.e. the constant that identifies the shape of the electrodes (see Fig. \ref{fig:comb_geometry}). The battery efficiency, reported in parenthesis, is computed as the ratio between the extracted charge at the end of the discharging and the theoretical capacity. }}
\label{fig:voltage2}
\end{figure}

\textit{The battery voltage} -  
Figure \ref{fig:voltage2} shows the simulated battery voltage for different number of combs under constant current regimes for 1.0 and 8.0 C-rates. As resulted in the former section, the operating battery voltage increases with $n$. The electrode morphology impacts on the amount of the extracted charge in a similar way: the higher the number of combs, the higher the efficiency of the battery. At 1.0 C-rate, the efficiency of the battery exceeds 90\% efficiency with 15 combs on both electrodes. The striking impact of the morphology on discharge rate is seen clearly also for high C-rate (equal to 8 in Fig. \ref{fig:voltage2}b).

\begin{figure}[htbp]
\centering
 \includegraphics[width=16.5cm]{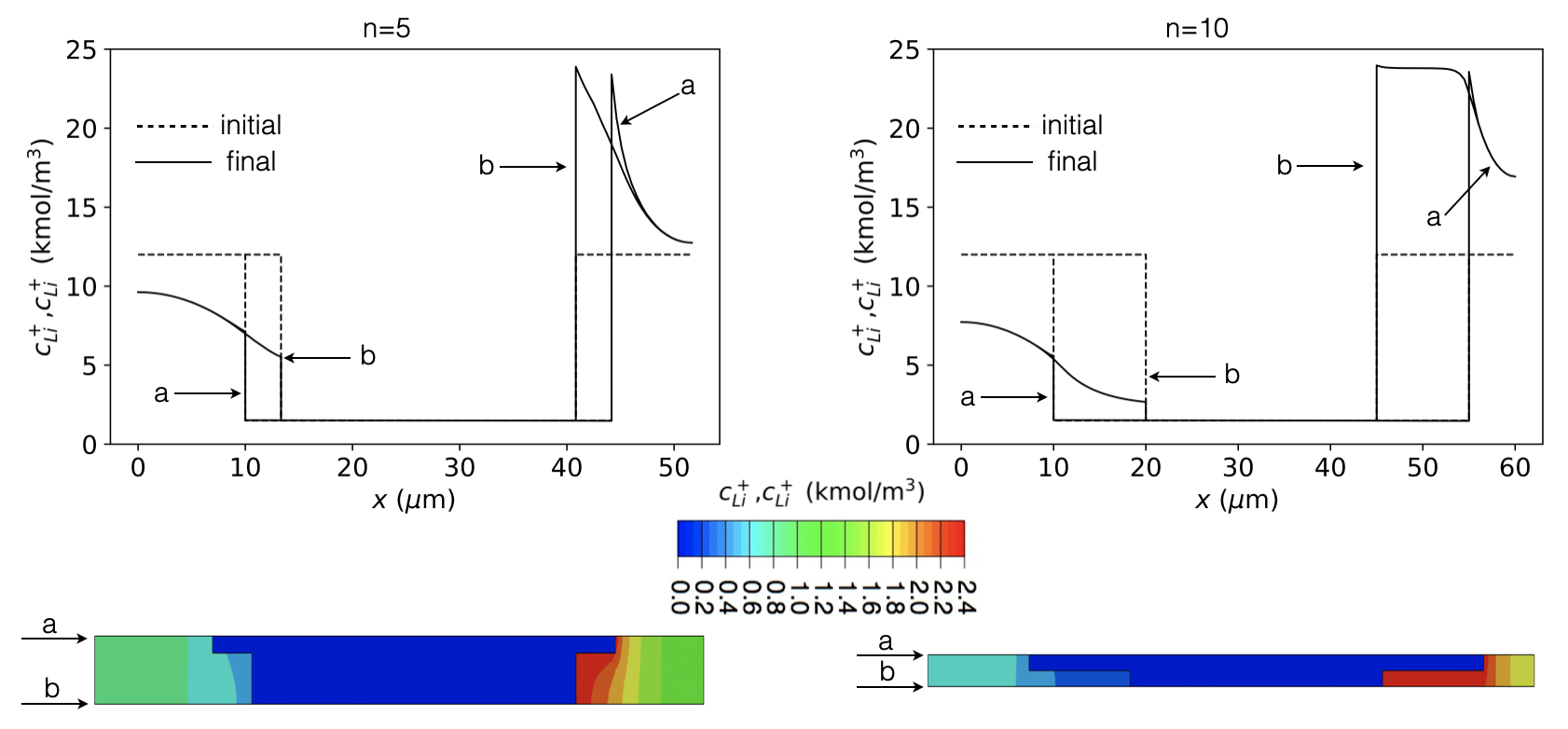}
\caption{\em{Comparison between the lithium distribution in the battery at the end of 8.0 C-rate discharging process for battery morphologies $n=5$ and $n=10$.}}
\label{fig:cLi2}
\end{figure}

\bigskip
\textit{The chemo-mechanical response} - The role of the battery geometry on the spatial distribution of lithium concentration is depicted in Fig. \ref{fig:cLi2} at 8.0 C-rate. By increasing the surface area of the electrodes and, even more relevant, by narrowing the thickness of combs, the lithium concentration in the cathode results more uniform. For $n=1$, the limiting factor is the saturation of lithium in the cathode, that prevents further discharge of the battery. On the contrary, for $n=10$ the depletion of lithium at the anode and the saturation at cathode are concurrent at the top of the combs, which are basically filled with stored lithium. Saturation in the cathode takes place in vicinity of the cathode-electrolyte interface, see the B plot in Fig. \ref{fig:cLi2} for $n=10$. This becomes the new cut-off factor of the battery and limits efficiency to 83.4\%, as seen in Fig. \ref{fig:voltage2}b. Efficiency increases with $n$, but the size of electrodes becomes higher and higher, see Fig. \ref{fig:n_comparison}b up to becoming unrealistic. Most likely, a three dimensional design of the electrode may relieve this drawback and allow optimal performance of the battery even at higher C-rates. This analysis will be the subject of further research.

\begin{figure}[htbp]
\centering
    \begin{subfigure}[!htb]{.49\textwidth}
        \centering
        \includegraphics[width=\textwidth]{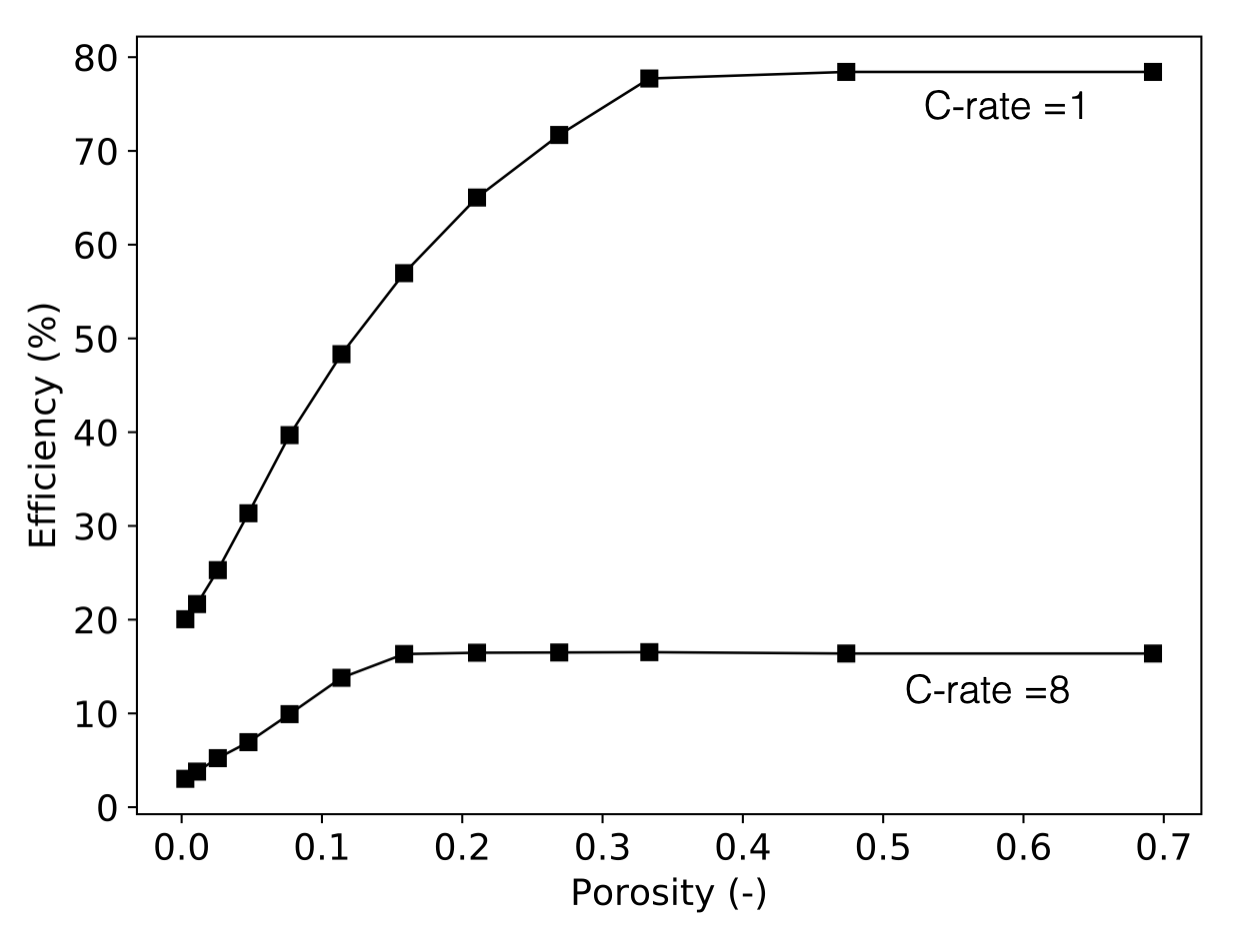}
        \caption{ }
	\label{fig:eff_por_a}
    \end{subfigure} 
    \begin{subfigure}[!htb]{.49\textwidth}
        \centering
        \includegraphics[width=\textwidth]{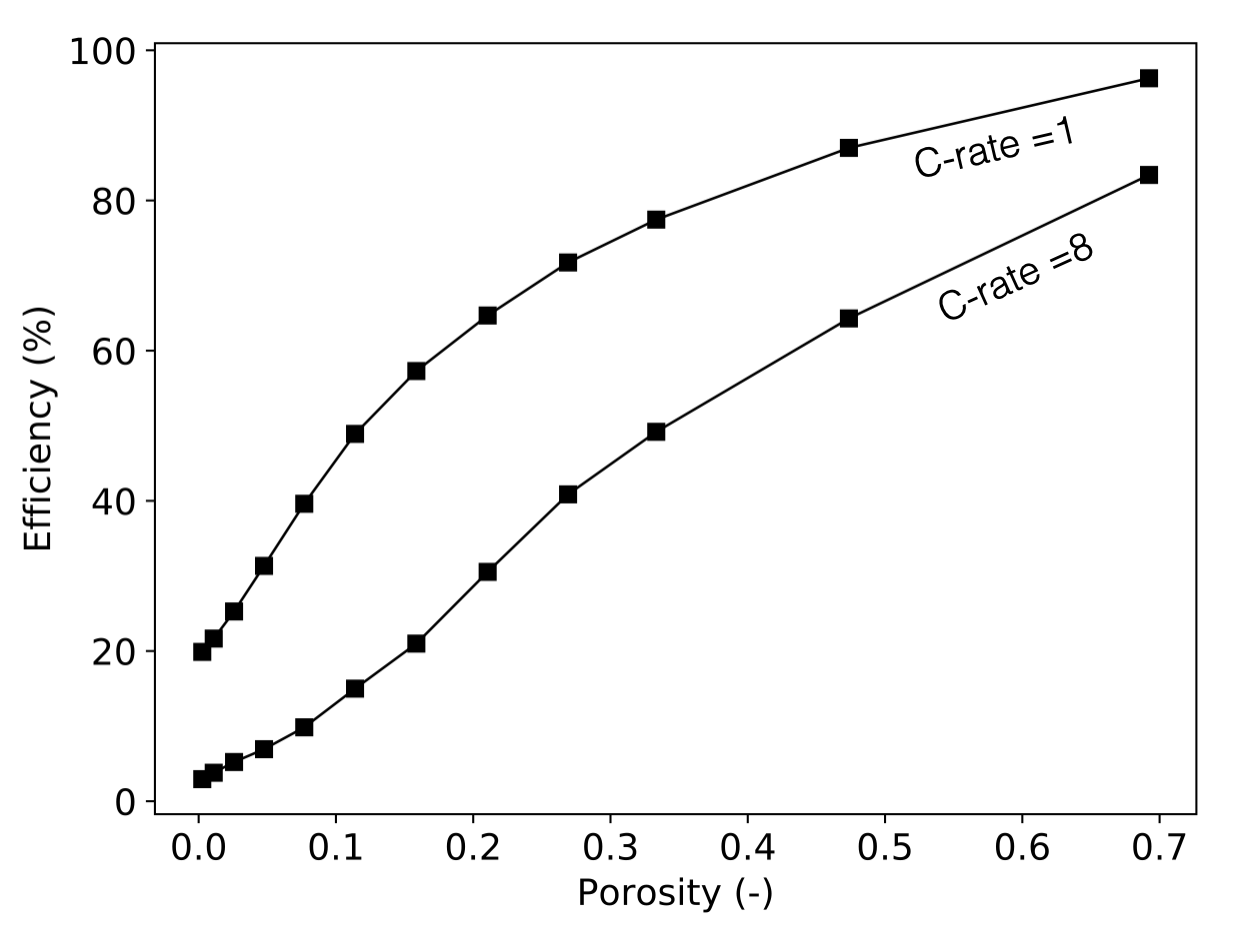}
        \caption{ }
	\label{fig:eff_por_b}
    \end{subfigure} 
\caption{\em{Influence of porosity on the efficiency, i.e. the ratio between stored charge and the cathodic theoretical capacity. }}
\label{fig:eff_por}
\end{figure}

\bigskip
The battery response as a function of porosity, already extensively discussed, is condensed in Fig. \ref{fig:eff_por}. Note that the same microstructure is used for both electrodes, hence a single parameter captures the porosity of the system. Figure \ref{fig:eff_por_a} depicts battery efficiency vs cathode porosity, whereas fig. \ref{fig:eff_por_b} refers to both electrodes.

\section{Conclusions}
In this study, the effect of the architecture of the electrode in Li-ion batteries has been investigated via thermo-chemo-mechanically coupled modeling and simulations. Insightful conclusions can be drawn, for which the battery efficiency is largely amplified by a proper design of the microstruture of the electrodes. At high C-rates, the battery cut off is caused by specific limiting factors: in discharge, as it emerged from the simulations, they stand in the concentration of inserted lithium in the cathode at the electrode/electrolyte interface, which prevents further discharge and limits the battery efficiency to few percent at 8 C-rates. An optimal design allows overcoming those limiting factors, redistributing the lithium owing to two mechanisms, the widening of oxidation/reaction surface area and the narrowing of the thickness, which allow almost complete filling of the electrodes. 
Although this paper deals with liquid electrolyte, similar conclusions might in principle be inferred in solid electrolytes. Furthermore, the role of the electrode microstructure exacerbates in materials with very low intercalation diffusivity, as for $\rm LiFePO_4$. In fact, very low diffusivities cause lithium to accumulate near the electrolyte interfaces, a limiting factor that structured electrodes can overcome as shown in this paper.

We shall point out that in several respects our model, and thus our conclusions in terms of increment of battery efficiency, does not capture complete realism. 
Modelling of fade and power loss with electrochemical cycling has been a significant research branch in LIBs modelling \cite{gargh2021correlating,saxena2021role,AroraEtAlJES1998} that was not accounted for. The main mechanisms of aging can be categorized in four groups, namely: surface film formation (solid electrolyte interphase (SEI) \cite{Ekstrom01012015,LiEtaAlJES2014,PinsonBazant2013,Liu_2014,WangEtAlCM2018,Cheng2015,Christensen2004,Koerver2017,Nie2013,RejovitzkyEtAlJMPS2015,Shin2015a,Shin2015b,TangJES2012,Xie2014,GUO2020104257,LIU202099}, bulk changes (phase segregation), mechanical effects due to lithiation (fracturing, dissipation, grinding), and parasitic reactions (corrosion, binder degradation)\cite{Barre2013,xiong2020lithium,vseruga2021continuous}. 
Those mechanisms will alter the numerical evidence in the increment of efficiency that our simulations highlighted, but will not diminish the conceptual relevance of architected electrodes. For this sake, further research will be elaborated in the near future, to further validate the main outcomes of the present note in the presence of aging and degradation, accentuated by mechanical detrimental effects \cite{muller2019study,muller2019effects}.


\bibliographystyle{unsrt}
\bibliography{/Users/albertosalvadori/Bibliography/Bibliography}

\begin{thebibliography}{10}

\bibitem{harper2019recycling}
G.~Harper, R.~Sommerville, E.~Kendrick, L.~Driscoll, P.~Slater, R.~Stolkin,
  A.~Walton, P.~Christensen, O.~Heidrich, and S.~Lambert.
\newblock Recycling lithium-ion batteries from electric vehicles.
\newblock {\em NATURE}, 575(7781):75--86, 2019.

\bibitem{lain2019design}
M.J. Lain, J.~Brandon, and E.~Kendrick.
\newblock Design strategies for high power vs. high energy lithium ion cells.
\newblock {\em BATTERIES}, 5(4):64, 2019.

\bibitem{wang2018toward}
J.~Wang, Q.~Sun, X.~Gao, C.~Wang, W.~Li, F.B. Holness, M.~Zheng, R.~Li, A.D.
  Price, and X.~Sun.
\newblock Toward high areal energy and power density electrode for li-ion
  batteries via optimized 3d printing approach.
\newblock {\em ACS APPL MATER INTER}, 10(46):39794--39801, 2018.

\bibitem{kremer2020manufacturing}
L.S. Kremer, A.~Hoffmann, T.~Danner, S.~Hein, B.~Prifling, D.~Westhoff,
  C.~Dreer, A.~Latz, V.~Schmidt, and M.~Wohlfahrt-Mehrens.
\newblock Manufacturing process for improved ultra-thick cathodes in
  high-energy lithium-ion batteries.
\newblock {\em ENERGY TECHNOL-GER}, 8(2):1900167, 2020.

\bibitem{kuang2019thick}
Y.~Kuang, C.~Chen, D.~Kirsch, and L.~Hu.
\newblock Thick electrode batteries: principles, opportunities, and challenges.
\newblock {\em ADV ENERGY MATER}, 9(33):1901457, 2019.

\bibitem{mei2019effect}
W.~Mei, H.~Chen, J.~Sun, and Q.~Wang.
\newblock The effect of electrode design parameters on battery performance and
  optimization of electrode thickness based on the electrochemical--thermal
  coupling model.
\newblock {\em SUSTAIN ENERG FUELS}, 3(1):148--165, 2019.

\bibitem{AifBook}
K.E. Aifantis, S.A. Hackney, and R.V. Kumar.
\newblock {\em High energy density lithium batteries. Materials, Engineering,
  Applications}.
\newblock Wiley, 2010.

\bibitem{FrancoRCS2013}
A.A. Franco.
\newblock Multiscale modelling and numerical simulation of rechargeable
  {L}ithium ion batteries: concepts, methods and challenges.
\newblock {\em RSC ADVANCES}, 3(13027), 2013.

\bibitem{GrazioliEtAlCM2016}
D.~Grazioli, M.~Magri, and A.~Salvadori.
\newblock Computational modeling of {L}i-ion batteries.
\newblock {\em COMPUT MECH}, 58(6):889--909, 2016.

\bibitem{liu2019simultaneous}
B.~Liu, X.~Wang, H.S. Chen, S.~Chen, H.~Yang, J.~Xu, H.~Jiang, and D.N. Fang.
\newblock A simultaneous multiscale and multiphysics model and numerical
  implementation of a core-shell model for lithium-ion full-cell batteries.
\newblock {\em J APPL MECH}, 86(4), 2019.

\bibitem{arunachalam2019full}
H.~Arunachalam and S.~Onori.
\newblock Full homogenized macroscale model and pseudo-2-dimensional model for
  lithium-ion battery dynamics: comparative analysis, experimental verification
  and sensitivity analysis.
\newblock {\em J ELECTROCHEM SOC}, 166(8):A1380, 2019.

\bibitem{battiato2019theory}
I.~Battiato, D.~O'Malley, C.T. Miller, P.S. Takhar, F.J. Vald{\'e}s-Parada, and
  B.D. Wood.
\newblock Theory and applications of macroscale models in porous media.
\newblock {\em TRANSPORT POROUS MED}, 130(1):5--76, 2019.

\bibitem{higa2017comparing}
K.~Higa, S.L. Wu, D.Y. Parkinson, Y.~Fu, S.~Ferreira, V.~Battaglia, and
  V.~Srinivasan.
\newblock Comparing macroscale and microscale simulations of porous battery
  electrodes.
\newblock {\em J ELECTROCHEM SOC}, 164(11):E3473, 2017.

\bibitem{lu20203d}
X.~Lu, A.~Bertei, D.P. Finegan, C.~Tan, S.R. Daemi, J.S. Weaving, K.B. O'Regan,
  T.~Heenan, G.~Hinds, and E.~Kendrick.
\newblock {3D microstructure design of Lithium-ion battery electrodes assisted
  by X-ray nano-computed tomography and modelling}.
\newblock {\em NAT COMMUN}, 11(1):1--13, 2020.

\bibitem{kim2018multiphysics}
S.~Kim, J.~Wee, K.~Peters, and H.~Huang.
\newblock Multiphysics coupling in lithium-ion batteries with reconstructed
  porous microstructures.
\newblock {\em J PHYS CHEM C}, 122(10):5280--5290, 2018.

\bibitem{zhu2021numerical}
X.~Zhu, Y.~Xie, H.~Chen, and W.~Luan.
\newblock Numerical analysis of the cyclic mechanical damage of li-ion battery
  electrode and experimental validation.
\newblock {\em INT J FATIGUE}, 142:105915, 2021.

\bibitem{sonwane2021coupling}
A.~Sonwane, C.~Yuan, and J.~Xu.
\newblock Coupling effect of state-of-charge and strain rate on the mechanical
  behavior of electrodes of 21700 lithium-ion battery.
\newblock {\em J ELECTROCHEM EN CONV STOR}, 18(2), 2021.

\bibitem{ZhangAEM2021}
X.~Zhang, Z.~Ju, Y.~Zhu, K.~J. Takeuchi, E.~S. Takeuchi, A.~C. Marschilok, and
  G.~Yu.
\newblock Multiscale understanding and architecture design of high energy/power
  lithium‐ion battery electrodes.
\newblock {\em ADV ENERGY MATER}, 11(2), 2021.

\bibitem{YanAEM2018}
C.~Yan, Y.~Zhu, Z.~Fang, C.~Lv, X.~Zhou, G.~Chen, and G.~Yu.
\newblock Heterogeneous molten salt design strategy toward coupling
  cobalt--cobalt oxide and carbon for efficient energy conversion and storage.
\newblock {\em ADV ENERGY MATER}, 8(23), 2018.

\bibitem{JiNL2012}
H.~Ji, L.~Zhang, M.~T. Pettes, H.~Li, S.~Chen, L.~Shi, R.~Piner, and R.~S.
  Ruoff.
\newblock Ultrathin graphite foam: A three-dimensional conductive network for
  battery electrodes.
\newblock {\em NANO LETT}, 12(5):2446--2451, 2012.

\bibitem{AguiloRSC2016}
N.~Aguil{\`o}-Aguayo, P.~Pena~Espi{\~a}eira, A.~P.~Manian, and T.~Bechtold.
\newblock Three-dimensional embroidered current collectors for ultra-thick
  electrodes in batteries.
\newblock {\em RSC ADVANCES}, 6(74):69685--69690, 2016.

\bibitem{EvanoffAM2012}
K.~Evanoff, J.~Khan, A.~A. Balandin, A.~Magasinski, W.~J. Ready, T.~F. Fuller,
  and G.~Yushin.
\newblock Towards ultrathick battery electrodes: Aligned carbon nanotube --
  enabled architecture.
\newblock {\em ADV MATER}, 24(4):533--537, 2012.

\bibitem{SanderNE2016}
J.~S. Sander, R.~M. Erb, L.~Li, A.~Gurijala, and Y.~M. Chiang.
\newblock High-performance battery electrodes via magnetic templating.
\newblock {\em NAT ENERGY}, 1(8):1--7, 2016.

\bibitem{Park2010}
M.~Park, X.~Zhang, M.~Chung, G.B. Less, and A.M. Sastry.
\newblock A review of conduction phenomena in {L}i-ion batteries.
\newblock {\em J POWER SOURCES}, 195(24):7904--7929, 2010.

\bibitem{MangangJPS2016}
M.~Mangang, H.~J. Seifert, and W.~Pfleging.
\newblock Influence of laser pulse duration on the electrochemical performance
  of laser structured $lifepo_4$ composite electrodes.
\newblock {\em J POWER SOURCES}, 304(24-32), 2016.

\bibitem{Pfleging2018}
W.~Pfleging.
\newblock A review of laser electrode processing for development and
  manufacturing of lithium-ion batteries.
\newblock {\em Nanophotonics}, 7(3):549--573, 2018.

\bibitem{TsudaEA2019}
T.~Tsuda, N.~Ando, S.~Nakamura, Y.~Ishihara, N.~Hayashi, N.~Soma, T.~Gunji,
  T.~Tanabe, T.~Ohsaka, and F.~Matsumoto.
\newblock Improvement of high-rate discharging performance of $lifepo_4$
  cathodes by forming micrometer-sized through-holed electrode structures with
  a pico-second pulsed laser.
\newblock {\em ELECTROCHIM ACTA}, 296:27--38, 2019.

\bibitem{ParkJIEC2019}
J.~Park, S.~Hyeon, S.~Jeong, and H.-J. Kim.
\newblock Performance enhancement of li-ion battery by laser structuring of
  thick electrode with low porosity.
\newblock {\em J IND ENG CHEM}, 70:178--185, 2019.

\bibitem{LimJES2014}
D.~G. Lim, D.-W. Chung, R.~Kohler, J.~Proell, C.~Scherr, W.~Pfleging, and R.~E.
  Garc{\'\i}a.
\newblock Designing 3d conical-shaped lithium-ion microelectrodes.
\newblock {\em J ELECTROCHEM SOC}, 161(3):A302--A307, 2014.

\bibitem{SalvadoriEtAlJPS2015}
A.~Salvadori, D.~Grazioli, M.G.D. Geers, D.~Danilov, and P.H.L Notten.
\newblock A multiscale-compatible approach in modeling ionic transport in the
  electrolyte of (lithium ion) batteries.
\newblock {\em J POWER SOURCES}, 293:892--911, 2015.

\bibitem{NewmanInScrosatiBook2002}
K.E. Thomas, J.~Newman, and R.M. Darling.
\newblock Matematical modeling of {L}ithium batteries.
\newblock In W.~van Schalkwijk and B.~Scrosati, editors, {\em Advances in
  {L}ithium-ion batteries}. Kluwer Academic, New York, 2002.

\bibitem{SalvadoriEtAlIJSS2015}
A.~Salvadori, D.~Grazioli, and M.G.D. Geers.
\newblock Governing equations for a two-scale analysis of {L}i-ion battery
  cells.
\newblock {\em INT J SOLIDS STRUCT}, 59:90--109, 2015.

\bibitem{SalvadoriEtAlJMPS2013}
A.~Salvadori, E.~Bosco, and D.~Grazioli.
\newblock A computational homogenization approach for {L}i-ion battery cells.
  {P}art 1 - {F}ormulation.
\newblock {\em J MECH PHYS SOLIDS}, 65:114--137, 2014.

\bibitem{SalvadoriGrazioliBookCh2015}
A.~Salvadori and D.~Grazioli.
\newblock Computer simulation for battery design and lifetime prediction.
\newblock In B.~Scrosati, J.~Garche, and W.~Tillmetz, editors, {\em Advances in
  battery technologies for electric vehicles}, pages 417--442. Woodhead
  Publishing, 2015.

\bibitem{SalvadoriEtAlJPS2015b}
A.~Salvadori, D.~Grazioli, M.~Magri, M.G.D. Geers, D.~Danilov, and P.H.L.
  Notten.
\newblock On the role of saturation in modeling ionic transport in the
  electrolyte of ({L}i-ion) batteries.
\newblock {\em J POWER SOURCES}, 294:696--710, 2015.

\bibitem{Danilovetal2011}
D.~Danilov, R.A.H. Niessen, and P.H.L. Notten.
\newblock Modeling all-solid-state {L}i-ion batteries.
\newblock {\em J ELECTROCHEM SOC}, 158(3):A215--A222, 2011.

\bibitem{SalvadoriEtAlJMPS2018}
A.~Salvadori, R.M. McMeeking, D.~Grazioli, and M.~Magri.
\newblock A coupled model of transport-reaction-mechanics with trapping. {P}art
  {I} - small strain analysis.
\newblock {\em J MECH PHYS SOLIDS}, 114:1--30, 2018.

\bibitem{Ganser2019}
M.~Ganser, F.E. Hildebrand, M.~Kamlah, and R.M. McMeeking.
\newblock A finite strain electro-chemo-mechanical theory for ion transport
  with application to binary solid electrolytes.
\newblock {\em J MECH PHYS SOLIDS}, 125:681--713, 2019.

\bibitem{dileoetalJMPS2014}
C.~Di~Leo, E.~Rejovitzky, and L.~Anand.
\newblock A {C}ahn-{H}illiard-type phase-field theory for species diffusion
  coupled with large elastic deformations: Application to phase-separating
  {L}i-ion electrode materials.
\newblock {\em J MECH PHYS SOLIDS}, 70:1--29, 2014.

\bibitem{DiLeoIJSS2015a}
C.V. Di~Leo, E.~Rejovitzky, and L.~Anand.
\newblock Diffusion-deformation theory for amorphous silicon anodes: the role
  of plastic deformation on elecrochemical performance.
\newblock {\em INT J SOLIDS STRUCT}, 67-68:283--296, 2015.

\bibitem{C6CP04142F}
W.~Dreyer, C.~Guhlke, and R.~Muller.
\newblock A new perspective on the electron transfer: recovering the
  butler-volmer equation in non-equilibrium thermodynamics.
\newblock {\em PHYS CHEM CHEM PHYS}, 18:24966--24983, 2016.

\bibitem{DanilovNottenEA2020}
L.H.J. Raijmakers, D.L. Danilov, R.A. Eichel, and P.H.L. Notten.
\newblock An advanced all-solid-state li-ion battery model.
\newblock {\em ELECTROCHIM ACTA}, 330(135147), 2020.

\bibitem{C5CP03836G}
W.~Dreyer and R.~Guhlke, C.and~Muller.
\newblock Modeling of electrochemical double layers in thermodynamic
  non-equilibrium.
\newblock {\em PHYS CHEM CHEM PHYS}, 17:27176--27194, 2015.

\bibitem{malaveEA2014b}
V.~Malave, J.~Berger, H.~Zhu, and R.J. Kee.
\newblock A computational model of the mechanical behavior within reconstructed
  {L}i$_x${C}o{O}$_2$ {L}i-ion battery cathode particles.
\newblock {\em ELECTROCHIM ACTA}, 130:707--717, 2014.

\bibitem{PurkayasthaMcMeekingCM2012}
R.T. Purkayastha and R.M. McMeeking.
\newblock An integrated 2-{D} model of a {L}ithium ion battery: the effect of
  material parameters and morphology on storage particle stress.
\newblock {\em COMPUT MECH}, 50:209--227, 2012.

\bibitem{DanilovNottenEA2008}
D.~Danilov and P.H.L. Notten.
\newblock Mathematical modeling of ionic transport in the electrolyte of
  {L}i-ion batteries.
\newblock {\em ELECTROCHIM ACTA}, 53:5569--5578, 2008.

\bibitem{Reimers01081992}
Jan~N. Reimers and J.~R. Dahn.
\newblock Electrochemical and in situ {X}-ray diffraction studies of lithium
  intercalation in $\rm {L}i_x{C}o{O}_2$.
\newblock {\em J ELECTROCHEM SOC}, 139(8):2091--2097, 1992.

\bibitem{Bohn2013}
E.~Bohn, T.~Eckl, M.~Kamlah, and R.~McMeeking.
\newblock A model for {L}ithium diffusion and stress generation in an
  intercalation storage particle with phase change.
\newblock {\em J ELECTROCHEM SOC}, 160(10):A1638--A1652, 2013.

\bibitem{Mukhopadhyay2014}
A.~Mukhopadhyay and B.V. Sheldon.
\newblock Deformation and stress in electrode materials for {L}i-ion batteries.
\newblock {\em PROG MATER SCI}, 63:58--116, 2014.

\bibitem{RenganathanEtAl2010}
S.~Renganathan, G.~Sikha, S.~Santhanagopalan, and R.~E. White.
\newblock Theoretical analysis of stresses in a {L}ithium ion cell.
\newblock {\em J ELECTROCHEM SOC}, 157:155--163, 2010.

\bibitem{ChenJAPS2018}
J.~Chen, H.~Hu, S.~Li, and Y.~He.
\newblock Evolution of mechanical properties of polypropylene separator in
  liquid electrolytes for lithium ion batteries.
\newblock {\em J APPL POLYM SCI}, 135(27):46441, 2018.

\bibitem{Guo01102011}
M.~Guo and R.E. White.
\newblock Thermal model for lithium ion battery pack with mixed parallel and
  series configuration.
\newblock {\em J ELECTROCHEM SOC}, 158(10):A1166--A1176, 2011.

\bibitem{GeersCohesive2008}
M.J. Van~den Bosch, P.J.G. Schreurs, and M.G.D. Geers.
\newblock On the development of a 3d cohesive zone element in the presence of
  large deformations.
\newblock {\em COMPUT MECH}, 42:171--180, 2008.

\bibitem{ORTIZPANDOLFI}
M.~Ortiz and A.~Pandolfi.
\newblock Finite-deformation irreversible cohesive elements for
  three-dimensional crack-propagation analysis.
\newblock {\em INT J NUMER METH ENG}, 44:1267--1282, 1999.

\bibitem{MULTIZONE}
A.~Salvadori.
\newblock A symmetric boundary integral formulation for cohesive interface
  problems.
\newblock {\em COMPUT MECH}, 32(4-6):381--391, 2003.

\bibitem{PARK2012239}
K.~Park and G.H. Paulino.
\newblock Computational implementation of the ppr potential-based cohesive
  model in abaqus: Educational perspective.
\newblock {\em ENG FRACT MECH}, 93(Supplement C):239--262, 2012.

\bibitem{gmshreference}
C.~Geuzaine and J.-F. Remacle.
\newblock Gmsh: a three-dimensional finite element mesh generator with built-in
  pre- and post-processing facilities.
\newblock {\em INT J NUMER METH ENG}, 79(11):1309--1331, 2009.

\bibitem{lyu2021overview}
Y.~Lyu, X.~Wu, K.~Wang, Z.~Feng, T.~Cheng, Y.~Liu, M.~Wang, R.~Chen, L.~Xu, and
  J.~Zhou.
\newblock An overview on the advances of {LiCoO2} cathodes for lithium-ion
  batteries.
\newblock {\em ADV ENERGY MATER}, 11(2):2000982, 2021.

\bibitem{gargh2021correlating}
P.~Gargh, A.~Sarkar, Y.H. Lui, S.~Shen, C.~Hu, S.~Hu, I.C. Nlebedim, and
  P.~Shrotriya.
\newblock Correlating capacity fade with film resistance loss in fast charging
  of lithium-ion battery.
\newblock {\em J POWER SOURCES}, 485:229360, 2021.

\bibitem{saxena2021role}
S.~Saxena, Y.~Ning, R.~Thompson, and M.~Pecht.
\newblock Role of the rest period in capacity fade of {Graphite/LiCoO2}
  batteries.
\newblock {\em J POWER SOURCES}, 484:229246, 2021.

\bibitem{AroraEtAlJES1998}
P.~Arora, R.E. White, and M.~Doyle.
\newblock Capacity fade mechanisms and side reactions in lithium-ion batteries.
\newblock {\em J ELECTROCHEM SOC}, 145(10):3647--3667, 1998.

\bibitem{Ekstrom01012015}
H.~Ekstr{\"o}m and G.~Lindbergh.
\newblock A model for predicting capacity fade due to {SEI} formation in a
  commercial graphite/{L}i{F}e{PO}$_4$ cell.
\newblock {\em J ELECTROCHEM SOC}, 162(6):A1003--A1007, 2015.

\bibitem{LiEtaAlJES2014}
D.~Li, D.~Danilov, Z.~Zhang, H.~Chen, Y.~Yang, and P.H.L Notten.
\newblock Modeling the {SEI}-formation on graphite electrodes in
  {L}i{F}e{PO}$_4$ batteries.
\newblock {\em J ELECTROCHEM SOC}, 162(6):A858--A869, 2015.

\bibitem{PinsonBazant2013}
M.B. Pinson and M.Z. Bazant.
\newblock Theory of {SEI} formation in rechargeable batteries: Capacity fade,
  accelerated aging and lifetime prediction.
\newblock {\em J ELECTROCHEM SOC}, 160(2):A243--A250, 2013.

\bibitem{Liu_2014}
L.~Liu and M.~Zhu.
\newblock Modeling of {SEI} layer growth and electrochemical impedance
  spectroscopy response using a thermal-electrochemical model of li-ion
  batteries.
\newblock {\em {ECS} TRANS}, 61(27):43--61, oct 2014.

\bibitem{WangEtAlCM2018}
A.~Wang, S.~Kadam, H.~Li, S.~Shi, and Y.~Qi.
\newblock Review on modeling of the anode solid electrolyte interphase (sei)
  for lithium-ion batteries.
\newblock {\em NPJ COMP MAT}, 4(1):15, 2018.

\bibitem{Cheng2015}
X.B. Cheng, R.~Zhang, C.Z. Zhao, F.~Wei, J.G Zhang, and Q.~Zhang.
\newblock A review of solid electrolyte interphases on {L}ithium metal anode.
\newblock {\em ADV SCI}, page 1500213, 2015.

\bibitem{Christensen2004}
J.~Christensen and J.~Newman.
\newblock A mathematical model for the {L}ithium-ion negative electrode solid
  electrolyte interphase.
\newblock {\em J ELECTROCHEM SOC}, 151(11):A1977--A1988, 2004.

\bibitem{Koerver2017}
R.~Koerver, I.~Ayg{\"u}n, T.~Leichtwei{\ss}, C.~Dietrich, W.~Zhang, J.O.
  Binder, P.~Hartmann, W.G. Zeier, and J.~Janek.
\newblock Capacity fade in solid-state batteries: Interphase formation and
  chemomechanical processes in nickel-rich layered oxide cathodes and lithium
  thiophosphate solid electrolytes.
\newblock {\em CHEM MATER}, 29:5574−5582, 2017.

\bibitem{Nie2013}
M.~Nie, D.~Chalasani, D.P. Abraham, Y.~Chen, A.~Bose, and B.L. Lucht.
\newblock {L}ithium ion battery graphite solid electrolyte interphase revealed
  by microscopy and spectroscopy.
\newblock {\em J PHYS CHEM-US}, 117:1257--1267, 2013.

\bibitem{RejovitzkyEtAlJMPS2015}
E.~Rejovitzky, C.V. Di~Leo, and L.~Anand.
\newblock A theory and a simulation capability for the growth of a solid
  electrolyte interphase layer at an anode particle in a {L}i-ion battery.
\newblock {\em J MECH PHYS SOLIDS}, 78(210-230), 2015.

\bibitem{Shin2015a}
H.~Shin, J.~Park, S.~Han, A.M. Sastry, and W.~Lu.
\newblock Component-/structure-dependent elasticity of solid electrolyte
  interphase layer in {L}i-ion batteries: Experimental and computational
  studies.
\newblock {\em J POWER SOURCES}, 277:169--179, 2015.

\bibitem{Shin2015b}
H.~Shin, J.~Park, A.M. Sastry, and W.~Lu.
\newblock Degradation of the solid electrolyte interphase induced by the
  deposition of manganese ions.
\newblock {\em J POWER SOURCES}, 284:416--427, 2015.

\bibitem{TangJES2012}
M.~Tang, S.~Lu, and J.~Newman.
\newblock Experimental and theoretical investigation of
  solid-electrolyte-interphase formation mechanisms on glassy carbon.
\newblock {\em J ELECTROCHEM SOC}, 159(11):A1775--A1785, 2012.

\bibitem{Xie2014}
Y.~Xie, J.~Li, and C.~Yuan.
\newblock Multiphysics modeling of {L}ithium ion battery capacity fading
  process with solid-electrolyte interphase growth by elementary reaction
  kinetics.
\newblock {\em J POWER SOURCES}, 248:172--179, 2014.

\bibitem{GUO2020104257}
K.~Guo, R.~Kumar, X.~Xiao, B.V. Sheldon, and H.~Gao.
\newblock Failure progression in the solid electrolyte interphase (sei) on
  silicon electrodes.
\newblock {\em NANO ENERGY}, 68:104257, 2020.

\bibitem{LIU202099}
Y.~Liu, K.~Guo, C.~Wang, J.~Han, and H.~Gao.
\newblock Concentration dependent properties and plastic deformation facilitate
  instability of the solid-electrolyte interphase in li-ion batteries.
\newblock {\em INT J SOLIDS STRUCT}, 198:99--109, 2020.

\bibitem{Barre2013}
A.~Barre, B.~Deguilhem, S.~Grolleau, M.~Gerard, F.~Suard, and D.~Riu.
\newblock A review on {L}ithium-ion battery ageing mechanisms and estimations.
\newblock {\em J POWER SOURCES}, 241:680--689, 2013.

\bibitem{xiong2020lithium}
R.~Xiong, Y.~Pan, W.~Shen, H.~Li, and F.~Sun.
\newblock Lithium-ion battery aging mechanisms and diagnosis method for
  automotive applications: Recent advances and perspectives.
\newblock {\em RENEW SUST ENERG REV}, 131:110048, 2020.

\bibitem{vseruga2021continuous}
D.~{\v{S}}eruga, A.~Gosar, C.A. Sweeney, J.~Jaguemont, J.~Van~Mierlo, and
  M.~Nagode.
\newblock Continuous modelling of cyclic ageing for lithium-ion batteries.
\newblock {\em ENERGY}, 215:119079, 2021.

\bibitem{muller2019study}
V.~M{\"u}ller, R.G. Scurtu, M.~Memm, M.A. Danzer, and M.~Wohlfahrt-Mehrens.
\newblock Study of the influence of mechanical pressure on the performance and
  aging of lithium-ion battery cells.
\newblock {\em J POWER SOURCES}, 440:227148, 2019.

\bibitem{muller2019effects}
V.~M{\"u}ller, R.G. Scurtu, K.~Richter, T.~Waldmann, M.~Memm, M.A. Danzer, and
  M.~Wohlfahrt-Mehrens.
\newblock Effects of mechanical compression on the aging and the expansion
  behavior of {Si/C-composite| NMC811 in different lithium-ion battery cell
  formats}.
\newblock {\em J ELECTROCHEM SOC}, 166(15):A3796, 2019.

\end{thebibliography}

\appendix

\section{Finite element implementation}
\label{app:FEM}

\subsection{Non-dimensional governing equations and weak form}

{\bf{Intercalating electrodes}}  - 
Equations \eqref{eq:a_mass_balance} - \eqref{eq:steady_ch_cons} -  \eqref{eq:elct_bal_mom} have been rephrased in term adimensional variables

\begin{equation} \label{eq: nondim_var}
x_i^* = 	\frac{x_i}{\bar{l}} \, , \qquad t^* = \frac{t}{\bar{t}} \, ,  \qquad c_\text{Li}^* = \frac{c_\text{Li}}{\bar{c}} \, , \qquad \phi^* = \frac{\phi F}{R \, T}\, , \qquad u_i^* = \frac{u_i}{\bar{l}} \, , \qquad  \sigma_{ij}^* = \frac{\sigma_{ij}}{\bar{\sigma}} \, ,
\end{equation} 

\noindent
by introducing $\bar{l}$, $\bar{t}$, $\bar{c}$, $\bar{\sigma}$ as reference length, time, concentration, and stress respectively. In this way, the governing equations of intercalating electrodes are equivalent to the following non-dimensional ones

\begin{subequations} \label{eq: nondim_governing_elc}
\begin{align}
&\frac{ \partial c_{\text{Li}}^* }{ \partial t^*} \, + \text{div}^* [ \vect{h}_\text{Li}^*]= 0 \, , \label{eqch6:non_dim_massbalanceequationLss}
 \\ \nonumber
\\  
&\text{div}^* [ \vect{i}^*_a ] = 0  \, , \label{eqch6:non_dim_ohm}
\\ \nonumber
\\ 
&\text{div}^* [ \tensor{\sigma}_a^* ] = \vect{0} \, , \label{eqch6:non_dim_stressbalancess}
 \end{align}
 \end{subequations}

\noindent
where 

\begin{equation*}
\text{div}^* [ \vect{ h}_\text{Li}^*] =  \sum_{i=1}^2 \, \frac{ { h }_\text{Li}^*  }{ \partial x_i^*} \, , \qquad \text{div}^* [ \tensor{\sigma}_a^* ] =  \sum_{i=1}^2 \, \sum_{j=1}^2 \, \frac{ { \sigma}_{ij}^*  }{ \partial x_j^*} \, \vect{e}_i \, , \qquad  \vect{h}_\text{Li}^* = \frac{\vect{h}_\text{Li} \, \bar{t}}{ \bar{c} \, \bar{l}} \, , \qquad  \vect{i}_a^* = \frac{\vect{i}_a \, \bar{t}}{ \bar{c} \, \bar{l} \, F} \, .
\end{equation*}

\noindent
Note that equations \eqref{eq: nondim_governing_elc} have the same expression of \eqref{eq:a_mass_balance} - \eqref{eq:steady_ch_cons} -  \eqref{eq:elct_bal_mom} but are formulated in terms of non-dimensional variables and operators. Similarly, a non-dimensional counterpart of the constitutive laws \eqref{eq:stress_def} -  \eqref{eq:Li_flux} - \eqref{eq:ohmlaw} can be easily obtained as follow 

\begin{subequations}
\begin{gather*}
\tensor{\sigma}^*_a = K^*_{a} \,  \trace{\tensor{\varepsilon}_a - \tensor{\varepsilon}^{ch}} \mathds{1} + 2 \, G^*_a \, \deviatoric{\tensor{\varepsilon}_a} \, , \\
\\
\vect{h}^*_{\text{Li}} = -\diffusivity^*_\text{Li} \, \nabla^* \left[ {c^*_\text{Li}} \right]  + \frac{\diffusivity^*_\text{Li} \, \omega^*_{Li} }{ (R \, T)^* } \, c^*_\text{Li} \, \left( \frac{ \left(c_\text{Li}^{max} \right)^* - c^*_\text{Li} }{\left(c_\text{Li}^{max} \right)^*} \right) \, \gradient{\trace{\tensor{\sigma}^*_a}} \\
\\
\vect{i}^*_a = \kappa^*_a \, \nabla^* \left[ {\phi^*_a} \right] \, .
\end{gather*}
\end{subequations}

\noindent
with non-dimensional material parameters defined as

\begin{subequations} \nonumber
\begin{gather}
\diffusivity_\text{Li}^* = \frac{\diffusivity_\text{Li} \, \bar{t}}{\bar{l}^2} \, , \qquad \omega_\text{Li}^* = \omega_\text{Li} \, \bar{c} \, ,  \qquad ( R \, T )^* = R \, T \, \frac{\bar{c}}{\bar{\sigma}} \, , \qquad \left( c_\text{Li}^{max } \right)^*  = \frac{c_\text{Li}^{max}}{\bar{c}} \, , \\
\\  
 \qquad \kappa_a^* = \frac{\kappa_a \,  \bar{t} \, R \, T}{ \bar{c} \, \bar{l}^2 \, F^2} \, , \qquad K_a^* = \frac{K_a}{\bar{\sigma}} \, , \qquad  G_a^* = \frac{G_a}{\bar{\sigma}}  \, .
\end{gather}
\end{subequations}

\noindent
To include the effect of stress gradient in equation \eqref{eq:Li_flux}  we define an additional dimensionless variable $\Sigma^*$ as 

\begin{equation} \label{eq: additional_degree}
\Sigma^* - \trace{\tensor{\sigma}^* }   = 0 \, ,
\end{equation}

\noindent
which will be approximate as an explicit degree of freedom. Eq. \eqref{eq: additional_degree} is then added to the set of governing equations \eqref{eq: nondim_governing_elc} for the numerical resolution of the problem.

\medskip

{\bf{Electrolyte}}  - Following the same procedure adopted for electrodes, governing equations \eqref{eq:el_mass_balance} - \eqref{eq:el_maxwell} - \eqref{eq:el_mechanics} are first made dimensionless by introducing the following non dimensional variables

\begin{gather} \label{eq: nondim_var_el}
x_i^* = 	\frac{x_i}{\bar{l}} \, , \qquad t^* = \frac{t}{\bar{t}} \, ,  \qquad c_{\text{Li}^+}^* = \frac{c_{\text{Li}^+}}{\bar{c}} \, ,  \qquad c_{\text{X}^-}^* = \frac{c_{\text{X}^-}}{\bar{c}} \, ,\nonumber \\
\qquad \phi^* = \frac{\phi F}{R \, T}\, , \qquad u_i^* = \frac{u_i}{\bar{l}} \, , \qquad  \sigma_{ij}^* = \frac{\sigma_{ij}}{\bar{\sigma}} \, ,
\end{gather} 

\noindent
with $\bar{l}$, $\bar{t}$, $\bar{c}$, $\bar{\sigma}$ representing reference length, time, concentration, and stress respectively. Taking advantage of the definitions \eqref{eq: nondim_var_el}, the balance laws of electrolyte are equivalent to the following non-dimensional ones

\begin{subequations} 
\begin{align} \label{eq: nondim_governing_el}
 &\frac{ \partial c_{\text{Li}^+}^* }{ \partial t^*} \,   + \text{div}^* [ \vect{h}_{\text{Li}^+}^*]= 0 \, ,\\ \nonumber
\\ 
& \frac{ \partial c_{\text{X}^-}^* }{ \partial t^*} \,   + \text{div}^* [ \vect{h}_{\text{X}^-}^*]= 0 \, ,\\ \nonumber
\\  
&\text{div}^* \left[ \frac{\partial \vect{D}_e^*}{\partial t^*} + \left( \vect{h}_{\text{Li}^+}^* - \vect{h}_{\text{X}^-}^* \right) \right]= 0  \, ,\\ \nonumber
\\ 
&\text{div}^* [ \tensor{\sigma}_e^* ] = \vect{0} \, ,
 \end{align}
\end{subequations}

\noindent
where 

\begin{gather*}
\text{div}^* [ \vect{ h}_\beta^*] =  \sum_{i=1}^2 \, \frac{ { h }_\beta^*  }{ \partial x_i^*} \, , \quad \text{div}^* [ \tensor{\sigma}_e^* ] =  \sum_{i=1}^2 \, \sum_{j=1}^2 \, \frac{ { \sigma }_{ij}^*  }{ \partial x_j^*} \, \vect{e}_i \, , \quad  \vect{h}_\beta^* = \frac{\vect{h}_\beta \, \bar{t}}{ \bar{c} \, \bar{l}} \, , \quad \vect{D}_e^* = \frac{\vect{D}_e }{ \bar{c} \, \bar{l} \, F} \, , 
\end{gather*}

\noindent
and

\begin{gather*}
\qquad \beta= \text{Li}^+ , \, \text{X}^- \, .
\end{gather*}

\noindent
In this way the dimensionless form of the constitutive laws \eqref{eq:ionic_fluxes} - \eqref{eq:const_def_disp_el} - \eqref{eq:el_mechanics} result as follow 

\begin{subequations}
\begin{gather*}
\vect{h}^*_{\text{Li}^+} = - \diffusivity^*_{\text{Li}^+} \, \nabla \left[ {c^*_{\text{Li}^+}} \right]  - \diffusivity^*_{\text{Li}^+}  \, c^*_{\text{Li}^+} \, \left( 1 - 2 \, \frac{c^*_{\text{Li}^+}}{ \left(c^{max} \right)^* } \right) \, \nabla^* \left[ {\phi^*_e} \right] \, , \\
\nonumber\\
\vect{h}^*_{\text{X}^-} = - \diffusivity^*_{\text{X}^-} \, \nabla^* \left[ {c_{\text{X}^-}} \right] +\diffusivity^*_{\text{X}^-}  \, c^*_{\text{X}^-} \, \left( 1 - 2 \, \frac{c^*_{\text{X}^-}}{ \left(c^{max} \right)^*} \right) \, \nabla^* \left[ {\phi^*_e} \right]\, , \\
\nonumber\\
\vect{D}^* = - \permittivity^* \, \nabla^*  \left[  {\phi^*_e}\right] \, , \\
\nonumber \\
\tensor{\sigma}^*_e = K^*_{e} \,  \trace{\tensor{\varepsilon}_e} \mathds{1} + 2 \, G^*_e \, \deviatoric{\tensor{\varepsilon}_e} \, .
\end{gather*}
\end{subequations}

\noindent where

\begin{subequations} \nonumber
\begin{gather}
\diffusivity_{\text{Li}^+}^* = \frac{\diffusivity_{\text{Li}^+} \, \bar{t}}{\bar{l}^2} \, , \qquad \diffusivity_{\text{X}^-}^* = \frac{\diffusivity_{\text{X}^-} \, \bar{t}}{\bar{l}^2} \, , \qquad \left( \frac{F}{R \, T} \right)^* = 1 \, , \qquad \big( c^{max}  \big)^*  = \frac{c^{max}}{\bar{c}} \, , \\
\\
 \qquad \permittivity_e^* = \frac{\permittivity_e  \, R \, T}{ \bar{c} \, \bar{l}^2 \, F^2} \, , \qquad K_e^* = \frac{K_e}{\bar{\sigma}} \, , \qquad  G_e^* = \frac{G_e}{\bar{\sigma}}  \, .
\end{gather}
\end{subequations}

\medskip

{\bf{Interfaces}} - 
The interface conditions are applied for integrals defined on interfaces on $\Gamma_{an}$ and $\Gamma_{ca}$ for both electrodes and electrolyte. It is convenient to gather together all the contributions as follow;

\begin{equation}
\begin{aligned}
\hspace{0.5cm}  \int_{\Gamma_{an}} \hat{\Delta}c_{\text{Li}} \; h_{BV} \left( \Delta \phi \right) \, \text{d}s +  \int_{\Gamma_{an}} \hat{\Delta}\phi \; i_{BV} \left( \Delta \phi\right) \, \text{d}s \, + \,  \int_{\Gamma_{an}} \, \hat{\Delta} \vect{u} \,  \cdot \, \vect{T}_\Gamma \left( \Delta \vect{u} \right)  \,  \text{d}s \,  \\\hspace{10cm} 
 \end{aligned}
\end{equation}
and 
\begin{equation}
\begin{aligned}
 &  \,  \,   \int_{\Gamma_{ca}} \hat{\Delta}c_{\text{Li}} \; h_{BV} \left(  \Delta \phi \right) \, \text{d}s +  \int_{\Gamma_{ca}} \hat{\Delta}\phi \; i_{BV} \left( \Delta \phi\right) \, \text{d}s \, + \,  \int_{\Gamma_{ca}} \, \hat{\Delta} \vect{u} \,  \cdot \, \vect{T}_\Gamma \left( \Delta \vect{u} \right)  \,  \text{d}s \, ,
 \end{aligned}
\end{equation}

Symbol $\Delta$ defines the jump of a certain variable at the interfaces ( $\alpha$ stands for an  and ca depending on the interface $\Gamma$ where integrals are computed.) 

\medskip

{\bf{Weak form}} -
The weak formulation results from multiplying the strong form of governing equations  by a suitable set of tests functions and performing an integration upon the domain, exploiting the \textit{integration by parts} formula with the aim of reducing the order of differentiation in space. The asterisk is omitted for the sake of readability. In conclusion, the overall weak form of battery governing equations can be written in the time interval $ \left[ 0 , \,  t_f \right] $ as 

\medskip

$\text{Find} \, y (\vect{x}, t) \in \mathcal{V}^{ \left[ 0 , \, t_f  \right] } \, \text{such that} $

\begin{equation} \label{eq:weak_form}
 \frac{\partial }{\partial t} \, b \left( \hat{y} (\vect{x}) , \, c_L ( \vect{x} , t )  \right) + a \left(  \hat{y} (\vect{x}) , \, y ( \vect{x}, t ) \right)  = f \left( \hat{y} (\vect{x})  \right)  \qquad \forall \, \hat{y} (\vect{x}) \in \mathcal{V} \end{equation}

\medskip

\noindent
where

\begin{equation*}
\begin{aligned}
\hspace{0.5cm} b \left( \hat{y}(\vect{x}), \, c_L ( \vect{x}, t )  \right) \, &= \,  \int_{\Omega_a} \,  \hat{c}_L \, c_L  \, \text{d}A \,  + \,  \int_{\Omega_e} \,  \hat{c}_{\text{Li}^+} \, c_{\text{Li}^+}   \, \text{d}A \, 
 + \int_{\Omega_e} \,  \hat{c}_{\text{X}^-} \, c_{\text{X}^-}   \, \text{d}A \, + \,   \int_{\Omega_e} \, \permittivity_e \,  \nabla [ \, \hat{\phi}_e \, ] \cdot \gradient{\phi_e}  \, \text{d}A \,, \hspace{10cm}
 \end{aligned}
\end{equation*}

\begin{align*}
 \hspace{0.6cm} a \left(  \hat{y}(\vect{x}) , \, y ( \vect{x}, t ) \right)  \, &= \,  \int_{\Omega_a}  \gradient{\hat{c_L}} \cdot \bigg\{ \diffusivity_L \, \gradient{c_L} - \diffusivity_\Sigma \left( c_L \right) \gradient{\Sigma} \bigg\} \, \text{d}A  \, + \,  \int_{\Omega_a} \,  \kappa_a \,  \nabla [\, \hat{\phi}_a \,  ] \cdot    \gradient{\phi_a}  \, \text{d}A  \, + \hspace{10cm} \\
 & \, +\, \int_{\Omega_a} \,  \nabla_S \big[ \, \vect{\hat {u}}_a \,  \big]  :  \tensor{\sigma}_a ( c_L , \Sigma , \vect{u}_a) \,  \text{d} A \, + \, \int_{\Omega_a} \hat{\Sigma} \Big\{  \Sigma - \trace{\tensor{\sigma}_a ( c_L , \Sigma ,\vect{u}_a )} \Big\} \, \text{d}A   \, + \,  \\
 &  \, + \,   \int_{\Omega_e}  \gradient{\hat{c}_{\text{Li}^+}} \cdot \bigg\{ \diffusivity_{\text{Li}^+} \, \gradient{c_{\text{Li}^+}} + \diffusivity_\phi^+ \left( c_{\text{Li}^+} \right) \gradient{\phi_e} \bigg\} \, \text{d}A \, + \, \\
 &  \, + \,   \int_{\Omega_e}  \gradient{\hat{c}_{\text{X}^-}} \cdot \bigg\{ \diffusivity_{\text{X}^-} \, \gradient{c_{\text{X}^-}} - \diffusivity_\phi^- \left( c_{\text{X}^-} \right) \gradient{\phi_e} \bigg\} \, \text{d}A \, + \, \\
 &  \, + \, \int_{\Omega_e} \gradient{\hat{\phi}_e} \cdot \Bigg\{ \diffusivity_{\text{Li}^+}  \gradient{c_{\text{Li}^+}}  + \diffusivity_{\phi}^+ \left( c_{\text{Li}^+} \right) \,  \gradient{\phi_e} \, 
 -  \diffusivity_{\text{X}^-}  \gradient{c_{\text{X}^-}}  + \diffusivity_{\phi}^- \left( c_{\text{X}^-} \right)  \,  \gradient{\phi_e} \Bigg\}  \, \text{d}A  \, + \, \\
 &  \, + \, \int_{\Omega_e} \,  \nabla_S \big[ \, \vect{\hat {u}}_e \,  \big]  :  \tensor{\sigma}_e (\vect{u}_e) \,  \text{d} A \, + \, \\
 &  \, + \, \int_{\Gamma_{an}} \hat{\Delta}c_{\text{Li}} \; h_{BV} \left( \Delta \phi \right) \, \text{d}s +  \int_{\Gamma_{an}} \hat{\Delta}\phi \; i_{BV} \left( \Delta \phi\right) \, \text{d}s \, - \,  \int_{\Gamma_{an}} \, \hat{\Delta} \vect{u} \,  \cdot \, \vect{T}_\Gamma \left( \Delta \vect{u} \right)  \,  \text{d}s \, + \, \\
 &  \, + \,   \int_{\Gamma_{ca}} \hat{\Delta}c_{\text{Li}} \; h_{BV} \left(  \Delta \phi \right) \, \text{d}s +  \int_{\Gamma_{ca}} \hat{\Delta}\phi \; i_{BV} \left( \Delta \phi\right) \, \text{d}s \, - \,  \int_{\Gamma_{ca}} \, \hat{\Delta} \vect{u} \,  \cdot \, \vect{T}_\Gamma \left( \Delta \vect{u} \right)  \,  \text{d}s \, ,
\end{align*}

\begin{equation*}
\begin{aligned}
   \hspace{.5cm}f \left( \hat{y}(\vect{x}) \right)  \, =  \,  - \, & \int_{\partial^N \Omega_{an}} \hat{c}_L \, \left\{ \vect{h}_L \cdot \vect{n}_{an} \right\} \, \text{d}s \, - \,  \int_{\partial^N \Omega_{ca}} \hat{c}_L \, \left\{ \vect{h}_L \cdot \vect{n}_{ca} \right\} \, \text{d}s \, + \,  \hspace{10cm}\\
- \, &  \int_{\partial^N \Omega_{an}} \hat{\phi}_{an} \, \left\{ \vect{i} \cdot \vect{n}_{an} \right\} \, \text{d}s \,  - \,   \int_{\partial^N \Omega_{ca}} \hat{\phi}_{ca} \, \left\{ \vect{i} \cdot \vect{n}_{ca} \right\} \, \text{d}s \,  + \\
+ \,&  \int_{\partial^N \Omega_{an}} \, \vect{\hat{u}}_{an} \,  \cdot \, \left( \tensor{\sigma}_{an} \, \vect{n}_{an} \right) \,  \text{d}s  \, + \,   \int_{\partial^N \Omega_{ca}} \, \vect{\hat{u}}_{ca} \,  \cdot \, \left( \tensor{\sigma}_{ca} \, \vect{n}_{ca} \right) \,  \text{d}s \, \, .
\end{aligned}
\end{equation*}

\noindent
where $\diffusivity_\Sigma$, $\diffusivity_\phi^+$, $\diffusivity_\phi^-$ stands for

\begin{gather*}
\diffusivity_\Sigma =  \frac{ \diffusivity_L \, \omega_L}{R \, T} \,  c_L \, \left( \frac{c_L^{max} - c_L }{c_L^{max}} \right) \, , \quad
\diffusivity_\phi^+ =  \frac{ \diffusivity_{\text{Li}^+} \, F}{R \, T} \,  c_{\text{Li}^+} \, \left(1 -2 \, \frac{c_{\text{Li}^+} }{c^{max}} \right) \, , \quad
\diffusivity_\phi^- =  \frac{ \diffusivity_{\text{X}^-} \, F}{R \, T} \,  c_{\text{X}^-} \, \left(1 -2 \, \frac{c_{\text{X}^-}}{c^{max}} \right) \, , 
\end{gather*}

\noindent
and $\Omega_a = \Omega_{an} \cup \Omega_{ca}$ is the domain occupied by the electrodes. Symbol $\Delta$ defines the jump of a certain variable at the interfaces ($a$ stands for $an$ or $ca$ depending on the interface $\Gamma$ where integrals are computed) 

\begin{equation}
 \Delta c_{\text{Li}} = c_L - c_{\text{Li}^+} \, , \qquad \Delta \phi = \phi_a - \phi_{e} \, , \qquad  \Delta \vect{u} = \vect{u}_a - \vect{u}_e \, ,
\end{equation}

\noindent
while $\vect{T}_\Gamma$ is the traction normal to the interface

\begin{equation}
\vect{T}_\Gamma = \tensor{\sigma}_a \,  \vect{n}_a  = - \tensor{\sigma}_e \,  \vect{n}_e \, .
\end{equation}

$y = \left\{ c_L, \, \phi_a , \, \vect{u}_a, \, \Sigma , \, c_{\text{Li}^+}, \, c_{\text{X}^-}, \, \phi_e , \, \vect{u}_e \right\}$ collecting the time-dependent unknown fields. Column $\hat{y}$ collects the steady state test functions that correspond to the unknown fields in $y$, i.e. $\hat{y} =\{ $ $\hat{c}_L$, $\hat{\phi}_a$, $\vect{\hat{u}}_a$, $\hat{\Sigma}$, $\hat{c}_{\text{Li}^+}$, $\hat{c}_{\text{X}^-}$, $\hat{\phi}_e$, $\vect{\hat{u}}_e \}$. The identification of the functional space $\mathcal{V}$ falls beyond the scope of this work.

\end{document}